\pdfoutput=1
\documentclass[11pt]{article}
\usepackage[hmargin=2.54cm,vmargin=2.54cm]{geometry} % see geometry.pdf
\usepackage{geometry}
\usepackage{amsmath,amssymb,setspace,fancyhdr,geometry,url,color}
\usepackage{subcaption,graphicx,caption}
\usepackage[authoryear,round]{natbib}
\RequirePackage{lineno} 
\allowdisplaybreaks[0]
\allowbreak
\title{Improving Ice Sheet Model Calibration Using Paleoclimate and Modern Data}
\author{Won Chang, Murali Haran, Patrick Applegate, David Pollard}
\begin{document}
%\linenumbers
\newcommand{\Dir}{\mathrm{Dir}}
\newcommand{\ceil}[1]{\lceil #1 \rceil}
\newcommand{\thh}{^\mathrm{th}}
\newcommand{\modtwo}{\mathrm{[mod~2]}}
\newcommand{\thetaof}[2]{\theta \langle #1;#2\rangle}
\newcommand{\Mpa}{M_\mathrm{P,A}}
\newcommand{\Ma}{M_\mathrm{A}}
\newcommand{\rjaccept}{\mathcal{A}}

\newcommand{\matern}{Mat\'{e}rn }
\newcommand{\ba}{\ensuremath{\mathbf{a}}}
\newcommand{\bA}{\ensuremath{\mathbf{A}}}
\newcommand{\balpha}{\ensuremath{\boldsymbol{\alpha}}}
\newcommand{\barB}{\ensuremath{\mathbf{\bar{B}}}}
\newcommand{\barY}{\ensuremath{\bar{\mathbf{Y}}}}
\newcommand{\barZ}{\ensuremath{\mathbf{\bar{Z}}}}
\newcommand{\bb}{\ensuremath{\mathbf{b}}}
\newcommand{\bB}{\ensuremath{\mathbf{B}}}
\newcommand{\bC}{\ensuremath{\mathbf{C}}}
\newcommand{\bD}{\ensuremath{\mathbf{D}}}
\newcommand{\bV}{\ensuremath{\mathbf{V}}}
\newcommand{\bdelta}{\ensuremath{\boldsymbol{\delta}}}
\newcommand{\be}{\ensuremath{\mathbf{e}}}
\newcommand{\bepsilon}{\ensuremath{\boldsymbol{\epsilon}}}
\newcommand{\bG}{\ensuremath{\mathbf{G}}}
\newcommand{\bIn}{\ensuremath{\mathbf{I}_n}}
\newcommand{\bIN}{\ensuremath{\mathbf{I}_N}}
\newcommand{\bJ}{\ensuremath{\mathbf{J}}}
\newcommand{\bk}{\ensuremath{\mathbf{k}}}
\newcommand{\bK}{\ensuremath{\mathbf{K}}}
\newcommand{\bLambda}{\ensuremath{\boldsymbol{\Lambda}}}
\newcommand{\bM}{\ensuremath{\mathbf{M}}}
\newcommand{\bmu}{\ensuremath{\boldsymbol{\mu}}}
\newcommand{\bnu}{\boldsymbol{\nu}}
\newcommand{\blambda}{\boldsymbol{\lambda}}
\newcommand{\bolda}{\ensuremath{\mathbf{a}}}
\newcommand{\boldeta}{\ensuremath{\boldsymbol{\eta}}}
\newcommand{\bomega}{\ensuremath{\boldsymbol{\omega}}}
\newcommand{\bP}{\ensuremath{\mathbf{P}}}
\newcommand{\bphi}{\ensuremath{\boldsymbol{\phi}}}
\newcommand{\bpsi}{\ensuremath{\boldsymbol{\psi}}}
\newcommand{\bQ}{\ensuremath{\mathbf{Q}}}
\newcommand{\bR}{\ensuremath{\mathbf{R}}}
\newcommand{\bs}{\ensuremath{\mathbf{s}}}
\newcommand{\bSigma}{\ensuremath{\boldsymbol{\Sigma}}}
\newcommand{\btheta}{\ensuremath{\boldsymbol{\theta}}}
\newcommand{\bu}{\ensuremath{\mathbf{u}}}
\newcommand{\bh}{\ensuremath{\mathbf{h}}}
\newcommand{\bx}{\ensuremath{\mathbf{x}}}
\newcommand{\bv}{\ensuremath{\mathbf{v}}}
\newcommand{\bw}{\ensuremath{\mathbf{w}}}
\newcommand{\bW}{\ensuremath{\mathbf{W}}}
\newcommand{\bX}{\ensuremath{\mathbf{X}}}
\newcommand{\bxi}{\ensuremath{\boldsymbol{\xi}}}
\newcommand{\bY}{\ensuremath{\mathbf{Y}}}
\newcommand{\bz}{\ensuremath{\mathbf{z}}}
\newcommand{\bZ}{\ensuremath{\mathbf{Z}}}
\newcommand{\bzero}{\ensuremath{\mathbf{0}}}
\newcommand{\cl}{\ensuremath{c\ell}}
\newcommand{\Cov}{\ensuremath{\mbox{Cov}}}
\newcommand{\E}{\ensuremath{\mbox{E}}}
\newcommand{\Eta}{\ensuremath{\mbox{H}}}
\newcommand{\sigep}{\sigma_{\boldsymbol{\epsilon}}^2}
\newcommand{\Sigom}{\Sigma_{\boldsymbol{\omega}}}
\newcommand{\bU}{\ensuremath{\mathbf{U}}}
\newcommand{\pkg}[1]{{\normalfont\fontseries{b}\selectfont #1}}
\let\proglang=\textsf
\let\code=\texttt
\renewcommand{\thefootnote}{\fnsymbol{footnote}}

\maketitle
\doublespacing
%Understanding and projecting long-term ice volume change in WAIS is central to climate risk assessment and management. 
% ; this in turn results in poorly constrained future ice volume projections % over time scales that are relevant to the typical response time of WAIS to %warming climate (hundreds to thousands of years into the future)

\begin{abstract}
Human-induced climate change may cause significant ice volume loss from
the West Antarctic Ice Sheet (WAIS). Projections of ice volume change from
ice-sheet models and corresponding future sea-level rise have large
uncertainties due to poorly constrained input parameters. In most future
applications to date, model calibration has utilized only modern or recent
(decadal) observations, leaving input parameters that control the long-term
behavior of WAIS largely unconstrained. Many paleo-observations are in the
form of localized time series, while modern observations are non-Gaussian
spatial data; combining information across these types poses non-trivial
statistical challenges. Here we introduce a computationally
efficient calibration approach that utilizes both modern and
paleo-observations to generate better-constrained ice volume projections.
Using fast emulators built upon principal component analysis and a reduced
dimension calibration model, we can efficiently handle high-dimensional and
non-Gaussian data. We apply our calibration approach to the PSU3D-ICE model
which can realistically simulate long-term behavior of WAIS. Our results
show that using paleo observations in calibration significantly reduces
parametric uncertainty, resulting in sharper projections about the future
state of WAIS.
One benefit of using paleo observations is found to be that unrealistic
simulations with overshoots in past ice retreat and projected future regrowth
are eliminated.
\end{abstract}

\vspace*{.2in}

\section{Introduction}
Human-induced climate change may cause significant ice volume loss in the polar regions. The West Antarctic Ice Sheet (WAIS) is particularly vulnerable to the changing climate because much of the ice is grounded well below sea level. Previous studies suggest that ice volume loss in this area may result in up to 4 meters of sea level rise \citep{fretwell2013bedmap2}, which in turn will require significant resource allocations for adaptation and may therefore have a profound impact on human society \citep[e.g.][]{lempert2012characterizing}. 

Many recent modeling efforts have focused on simulating and projecting the past and future WAIS volume dynamics  \citep[e.g][]{bindschadler2013ice}.  The relevant time scale for many aspects of WAIS behavior is often in  hundreds to thousands of years;  this in turn necessitates simulating and projecting the evolution of WAIS for a very long time span \citep{ cornford2015century,feldmann2015collapse,golledge2015multi,gomez2015sea,ritz2015potential,winkelmann2015combustion}. 

In this work we use the PSU3D-ICE model \citep{pollard2009modelling,pollard2012simple,pollard2012description} to  simulate the long-term evolution of the West Antarctic Ice Sheet. Using a hybrid dynamical core that combines the shallow-ice and the shallow-shelf approximation, the model can realistically simulate the long-term behavior of the ice sheet with reasonable amount of computational effort. In contrast to higher-resolution models \citep[e.g.][]{gladstone2012calibrated,favier2014retreat,joughin2014marine} that are designed for relatively short simulations with more detailed physical processes, this modeling strategy enables us to simulate the long-term evolution of the West Antarctic Ice Sheet and utilize information from paleo-observations for model calibration. The approach here is a tradeoff between (i) reduced fidelity in
capturing details such as sills near modern grounding lines that may be
important for 10's-km scale retreat in the next hundred years, and (ii)
more robust calibration versus retreat over hundreds of kilometres and more pronounced
bedrock variations, which is arguably more relevant to larger-scale retreat
into the West Antarctic interior within the next several centuries.

Parametric uncertainty is an important source of uncertainty in projecting future WAIS volume change. Ice sheet models have input parameters that strongly affect the model behaviors; they are also poorly constrained  \citep{stone2010investigating,applegate2012assessment}. Various calibration methods have been proposed to reduce parametric uncertainty for Greenland \citep{chang2013ice,mcneall2013potential} and the Antarctica ice sheets model  \citep{gladstone2012calibrated,chang2015binary}. Although these recent studies have provided statistically sound ways of generating constrained future projections, they are mostly limited to generating short-term projections (i.e. a few hundred years from present) or utilizing modern or recent observations in the calibration. Inferring input parameters that are related to the long-term behavior of WAIS is crucial for generating well constrained projections in the relevant time scale (hundreds to thousands of years). Modern or recent observations often lack information on these parameters and therefore calibrating solely based on these information sources may result in poorly-constrained projections. {\color{black} Recent studies using heuristic approaches suggest that utilizing information from paleo data can reduce uncertainties in these long-term behavior related parameters  \citep{whitehouse2012deglacial,whitehouse2012new,briggs2013glacial,briggs2013evaluate,briggs2014data,golledge2014antarctic,maris2015model}.}

Here we propose an approach to simultaneously utilizing modern- and paleo-observations for ice sheet model calibration, generating  well-constrained future WAIS ice volume change projections. This work represents the first statistically rigorous approach for  calibrating an ice sheet model based on both modern and paleo data, and the resulting inference about parameters and projected sea level rise are therefore less uncertain than those obtained solely based on modern data  \citep[cf.][]{chang2015binary}. Our methodological contribution is to provide a computationally expedient approach to fuse information from modern and paleo data. Our dimension reduction methods build upon two different calibration approaches given by \citet{chang2013fast} and \citet{chang2015binary}, while also accounting for potential relationships between the two very different types of data -- the modern data are spatial and binary while the paleo data are in the form of a time series. A central contribution of this work is scientific. Based on our methods, we are able to show  explicitly how paleo data provides key new information about parameters of the ice sheet model, and we thereby show that utilizing paleo data in addition to modern ice sheet data virtually eliminates the possibility of zero (or negative) sea level rise.

The rest of the paper is organized as follows. Section \ref{section:ModelAndObs} introduces the model runs and the observational data sets used in our calibration experiment. Section \ref{section:DimReducedCalibration} explains our  computationally efficient reduced-dimension calibration approach that enables us to emulate and calibrate the PSU3D-ICE model using both the grounding line positions and the modern observations while avoiding the computational and inferential challenges. Section \ref{section:results} describes our calibration experiment results based on a simulated example and real observations. Finally, in Section \ref{section:discussion}, we summarize our findings and discuss caveats and possible improvements.

\section{Model Runs and Observational Data} \label{section:ModelAndObs}
In this section, we describe the ice-sheet model that we use to simulate  past and future West Antarctic Ice Sheet behavior, as well as the modern and paleo-data sets that we use to calibrate the ice-sheet model.

\subsection{Model Runs and Input Parameter Decription} \label{subsection:ModelDescription}

We calibrate the following four model parameters that are considered to be
important in determining the long-term evolution of the West Antarctic
Ice Sheet, yet whose values are particularly uncertain: the sub-ice-shelf oceanic
melt factor (OCFAC), the calving factor (CALV), the basal sliding coefficient
(CRH), and the asthenospheric relaxation \textit{e}-folding time (TAU).
OCFAC (non-dimensional) represents the strength of oceanic melting at the
base of floating ice shelves in response to changing ocean temperatures,
and CALV (non-dimensional) determines the rate of iceberg calving
at the outer edges of floating ice shelves. CRH (m year$^{-1}$ Pa$^{-2}$)
represents how slippery the bedrock is in areas around Antarctica that
are currently under ocean, i.e., how fast grounded ice slides over these areas
as it expands beyond the present extent. A higher value of CRH corresponds
to faster basal sliding and greater ice flux towards the ocean.
TAU (with units in thousands of years) represents the time scale for vertical bedrock displacements
in response to changing ice load. In this paper, we remap the parameter values
to the $[0,1]$ intervals for convenience. We refer to
\cite{chang2015binary} for more detailed description of these parameters.

We run the PSU3D-ICE model with 625 different parameter settings specified by a factorial design, with five different values for each parameter. Starting from 40,000 years before present, each model run is spun up until present and then projected 5,000 years into the future. For atmospheric forcing, we use the modern climatological Antarctic dataset from the Sea Rise project \citep{bindschadler2013ice} uniformly perturbed in proportion to a deep-sea-core d18O record \citep{pollard2009modelling,pollard2012description}. For oceanic forcing, we use the ocean temperature pattern  from AOGCM simulation runs generated by \cite{liu2009transient}. From each model run we extract the following two data sets and compare them to the corresponding observational record: (i) the time series of grounding line positions,  the location of the transition from grounded ice to ice shelf along the central flowline in the the Amundsen Sea Embayment (ASE) region (see Section \ref{subsection:obs_grounding_line} below for more details), and (ii) the modern binary spatial pattern of presence and absence of grounded ice in the ASE. The grounding line position time series (i) has 1,501 time points  from 15,000 years ago to the present, and the modern binary spatial pattern (ii) is a binary map with $86 \times 37$ pixels with a 20 km horizontal resolution. Because the time series of the grounding line position for each model run does not usually show much change until 15,000 years before present, we only use the time series after 15,000 years ago for our calibration. Note that each model output is in the form of high-dimensional multivariate data, which causes computational and inferential challenges described in Section \ref{subsubsection:challenges}. The corresponding  observational data sets, described below, have the same dimensionalities as the model output sets.

Note that, for some parameter settings, the past grounding line position shows unrealistically rapid retreat early in the time series and does not change for the rest of the time period. We have found that using these runs for building our emulator negatively affects the emulation performance for more realistic model outputs. Therefore, we have exclude the model runs that reach a grounding line position 500 meters inland from the modern grounding line before 10,000 years ago from our analysis and use the remaining 461 model runs for emulating the past grounding line position output.

\subsection{Paleo-records of Grounding Line Positions} \label{subsection:obs_grounding_line}
We take advantage of a very recent, comprehensive synthesis of Antarctic
grounding-line data since the last glacial maximum \citep{bentley2014community}. For the ASE sector, \cite{larter2014reconstruction} provide spatial maps of estimated grounding lines at 
5,000 year intervals from 25,000 years ago to the present. These maps are based primarily on many 
ship-based observations taken in the oceanic part of ASE using sonar 
(showing patterns of ocean-floor features formed by flow of past 
grounded ice) and shallow sediment cores (providing dating and information
on ice proximity, i.e., open ocean, ice proximal or grounded). There is considerable
uncertainty in the reconstructions, but general consensus for the
overall retreat in this sector, especially along the central flowline
of the major paleo-ice stream emerging from the confluence of Pine Island 
and Thwaites Glaciers and crossing the ASE \citep[cf. earlier 
synthesis by][]{kirshner2012post}. 

\subsection{Modern Observations}  \label{subsection:obs_modern}
We use a map of modern grounding lines deduced from the Bedmap2 dataset \citep{fretwell2013bedmap2}. Bedmap2 is the most recent all-Antarctic dataset that provides gridded maps of ice surface elevation, bedrock elevation, and ice thickness. These fields were
derived from a variety of sources, including satellite altimetry, airborne
and ground radar surveys, and seismic sounding. The nominal Bedmap2 grid
spacing is 1 km, but the actual coverage in some areas is sparser especially
for ice thickness. We deduce grounding line locations by a simple floatation
criterion at each model grid cell, after interpolating the
data to our coarser model grid. In this work, we use a part of the data that corresponds to the ASE sector, which is expected to be the largest contributor to future sea level rise caused by WAIS volume loss \citep{pritchard2012antarctic}. Since the observed modern binary pattern is derived from the highly accurate ice thickness measurements in the Bedmap2 data set and the model binary patterns are highly variable depending on the input parameter settings (see Section S1 and Figure S1 in the Supplement), the model outputs approximated by our emulator are accurate enough to provide a basis for calibration.

\section{Computer Model Emulation and Calibration Using Dimension Reduction} \label{section:DimReducedCalibration}

%As we discussed above, 
As explained in the preceding discussion, parameter inference is central to WAIS volume change projections; taking full advantage of all observational data, both paleo and modern, may result in reduced parametric uncertainty which in turn can result in well-constrained projections of volume change. In this section we describe our statistical approach for inferring input parameters for WAIS models while accounting for relevant sources of uncertainty. In Section \ref{subsection:GeneralFramework} we introduce our two-stage framework  \citep{bayarri2007computer,bhat2010computer,bhat2013inferring,chang2013fast} that consists of the emulation and the calibration steps: In the emulation step we build a Gaussian process emulator as a fast approximation to WAIS model outputs \citep{sacks1989design}. In the calibration step, we infer the input parameters for WAIS models by combining information from emulator output and observational data while accounting for the systematic model-observation discrepancy. The framework faces computational and inferential challenges when model output and observational data are high-dimensional data such as large spatial patterns or long time series, and the challenges are further exacerbated when the marginal distribution of model output and observational data cannot be modeled by a Gaussian distribution. Section \ref{subsection:RedDimApproaches} describes a computationally expedient reduced-dimension approach that mitigates these challenges for high-dimensional Gaussian and non-Gaussian data and enables us to utilize information from both past grounding line positions and modern binary patterns for calibration.

We use the following notation henceforth: Let $\btheta_1,\dots,\btheta_{q} \in R^4$ be the parameter settings at which we use both the past grounding line positions and the modern binary patterns  for emulation  and $\btheta_{q+1},\dots,\btheta_{p} \in R^4$ be the settings at which we use only the  the modern binary patterns (see Section \ref{subsection:ModelDescription} above for the reason why $q$ is less than $p$ in our experiment). We denote the past grounding line position time series from our WAIS model at a parameter setting $\btheta$ and a time point $t$ by $Y_1(\btheta,t)$. We let $\bY_1$ be a $q \times n$ matrix where its $i$th row is $[Y_1(\btheta_i,t_1),\dots,Y_1(\btheta_i,t_n)]$ and $t_1,\dots,t_n$ are time points at which the grounding line positions are recorded. We let $\bZ_1=[Z_1(t_1),\dots,Z_1(t_n)]$  be a vector of the observed time series of  past grounding line positions reconstructed from paleo records. Similarly, we denote the modern ice-no ice binary output at the parameter setting $\btheta$ and a spatial location $\bs$ by $Y_2(\btheta,\bs)$. We let $\bY_2$ be a $p \times m$ matrix where its $i$th row is $[Y_2(\btheta_i,\bs_1),\dots,Y_2(\btheta_i,\bs_m)]$ with model grid points $\bs_1,\dots,\bs_m$.   The corresponding observational data are denoted by an $m$-dimensional vector $\bZ_2=[Z_2(\bs_1),\dots,Z_2(\bs_m)]$. For our WAIS model emulation and calibration problem in Section \ref{section:results}, $n=1,501$, $m=3,182$, $p=625$, and $q=461$.

\subsection{Basic WAIS Model Emulation and Calibration Framework}\label{subsection:GeneralFramework}

In this subsection we explain the basic general framework for computer model emulation and calibration and describe the computational challenges posed by our use of high-dimensional model output and observational data. 

\subsubsection{Emulation and Calibration using Past Grounding Line Positions}
\label{subsubsection:GroundingEmulationCalibration}

We start with the model output $\bY_1$ and the observational data $\bZ_1$ for the past grounding line positions. Since computer model runs are available only at a limited number of parameter settings $q$, one needs to construct a statistical model for approximating the model output at a new parameter setting $\btheta$ by interpolating the existing model output obtained at the design points $\btheta_1,\dots,\btheta_q$ \citep{sacks1989design}. Constructing this statistical model requires us to build a Gaussian process that gives the following probability model for the existing $q$ model runs with $n$-dimensional output:
\begin{equation} \label{equation:GaussianEmul}
\mbox{vec}\left( \bY_1 \right) \sim N (\bX_1 \boldsymbol{\beta}_1, \bSigma (\bxi_1)),
\end{equation}
where $\mbox{vec}(.)$ is the vectorization operator that stacks the columns of a matrix into one column vector, and $\bX_1$ is a $nq\times b$ covariate matrix that contains all the time coordinates and the input parameters settings used in the $nq\times nq$ covariance matrix $\bSigma (\bxi_1)$ with a covariance parameter vector $\bxi_1$. The $b$-dimensional vector $\boldsymbol{\beta}_1$  contains all the coefficients for the columns of $\bX_1$. When the number of time points $n$ and the number of parameter settings $q$ are small, one can estimate the parameter $\bxi_1$ by maximizing the likelihood function corresponding to the probability model in \eqref{equation:GaussianEmul} and finding the conditional distribution of $Y_1(\btheta_i,t_1),\dots,Y_1(\btheta_i,t_n)$ given $\bY_1$ for any new value of $\btheta$ using the fitted Gaussian process with the maximum likelihood estimates $\hat{\boldsymbol{\beta}}_1$ and $\hat{\bxi}_1$. We call the fitted Gaussian process an emulator and denote the output at $\btheta$ interpolated by the emulator as $\boldeta(\btheta,\bY_1)$. Using the emulator, one can set up a model for inferring the input parameter $\btheta$ as follows \citep{kennedy2001bayesian,bayarri2007computer}:
\begin{equation} \label{equation:CalibModel}
\bZ_1=\boldeta(\btheta,\bY_1)+\bdelta,
\end{equation}
where $\bdelta$ is an $n$-dimensional random vector that represents model-observation discrepancy. The discrepancy term $\bdelta$ is often modeled by a Gaussian process that is independent of the emulated output $\boldeta(\btheta,\bY_1)$. Using a posterior density defined by the likelihood function that corresponds to the probability model in \eqref{equation:CalibModel} and a standard prior specification, one can estimate the input parameter $\btheta$ along with other parameters in the model via Markov Chain Monte Carlo (MCMC).  

\subsubsection{Emulation and Calibration using Modern Binary Observations} \label{subsubsection:BinaryEmulationCalibration}

Emulation and calibration for the modern ice-no ice binary patterns require additional consideration in model specification due to the binary nature of the data sets. Inspired by the generalized linear model framework, \cite{chang2015binary} specifies emulation and calibration models in terms of logits of model output and observational data. Let $\boldsymbol{\Gamma}=\left\{\gamma_{ij}\right\}$ be a $p \times m$-dimensional matrix whose element is the logit of the $(i,j)$th element in $\bY_2$, i.e., 
\begin{align*}
P(Y_2(\btheta_i,\bs_j)=y_{ij})=&\left(\frac{\exp(\gamma_{ij})}{1+\exp(\gamma_{ij})}\right)^{y_{ij}} \left(\frac{1}{1+\exp(\gamma_{ij})}\right)^{1-y_{ij}}\\
=&(1+\exp(-(2y_{ij}-1)\gamma_{ij}))^{-1},
\end{align*}
where the value of $y_{ij}$ is ether 0 or 1. Assuming conditional independence between the elements in $\bY_2$ given $\boldsymbol{\Gamma}$, one can use a Gaussian process that yields the probability model below to specify the dependence between the model output at different parameter settings and spatial locations,
\begin{equation} \label{equation:GaussianEmulGamma}
\mbox{vec}\left( \boldsymbol{\Gamma} \right) \sim N (\bX_2 \boldsymbol{\beta}_2, \bSigma (\bxi_2)),
\end{equation}
where the $mp\times c$ dimensional covariate matrix $\bX_2$, $c$-dimensional coefficient vector $\boldsymbol{\beta}_2$, and the $mp\times mp$ covariance matrix $\bSigma (\bxi_2)$ with a covariance parameter vector $\bxi_2$ are defined in the same way as in \eqref{equation:GaussianEmul}. One can find the maximum likelihood estimates $\hat{\boldsymbol{\beta}}_2$ and $\hat{\bxi}_2$ by maximizing the likelihood function corresponding to the probability model in \eqref{equation:GaussianEmulGamma}. The resulting Gaussian process emulator gives a vector of interpolated logits $\bpsi(\btheta,\bY_2)$ for a new input parameter value $\btheta$. 

The calibration model is also defined in terms of the logits of the observational data $\bZ_2$, denoted by an $m$-dimensional vector $\blambda=[\lambda_1,\dots,\lambda_m]$,
\begin{equation*}
\blambda=\bpsi(\btheta,\bY_2)+\bphi,
\end{equation*}
where $\bphi$ is an $m$-dimensional random vector representing the model-observation discrepancy defined in terms of the logits of $\bZ_2$. Again, assuming conditional independence between the elements in $\bZ_2$ given $\blambda$, the relationship between $\blambda$ and $Z_2(\bs_i)$ is given by
\begin{align} \label{equation:Z2andLambda}
\begin{split}
P(Z_2(\bs_j)=z_j)=&\left(\frac{\exp(\lambda_j)}{1+\exp(\lambda_j)}\right)^{z_j} \left(\frac{1}{1+\exp(\lambda_i)}\right)^{1-z_j}\\
=&(1+\exp(-(2z_j-1)\lambda_j))^{-1},
\end{split}
\end{align}
where $z_j$ takes a value of either 0 or 1. If the number of parameter settings $p$ and spatial locations $m$ are small, one can set up a posterior density based on the likelihood function corresponding to the probability models above and some standard prior specifications and might be able to infer $\btheta$ and other parameters using MCMC. 

\subsubsection{Combining Information from Two Data Sets in Calibration}
 
   We set up a calibration model to infer the input parameters in $\btheta$ based on the models described in Sections \ref{subsubsection:GroundingEmulationCalibration} and \ref{subsubsection:BinaryEmulationCalibration}. The main consideration here is how to model the dependence between $\bZ_1$ and $\bZ_2$, which can be translated to the dependence between the emulated outputs $\boldeta(\btheta,\bY_1)$ and $\boldsymbol{\psi}(\btheta,\bY_2)$ and the dependence between the discrepancy terms $\bdelta$ and $\bphi$. We model the dependence between  $\boldeta(\btheta,\bY_1)$ and $\boldsymbol{\psi}(\btheta,\bY_2)$ only through the input parameter $\btheta$ (i.e., we assume that $\boldeta(\btheta,\bY_1)$ and $\boldsymbol{\psi}(\btheta,\bY_2)$ are conditionally independent given the input parameter $\btheta$), because the emulators are independently constructed for $\bY_1$ and $\bY_2$. We do not introduce a conditional dependence between the emulators $\boldeta(\btheta,\bY_1)$ and $\boldsymbol{\psi}(\btheta,\bY_2)$ given $\btheta$ because the emulators are already highly accurate. The greater challenge is the dependence between the discrepancy terms $\bdelta$ and $\bphi$ because both terms are high-dimensional and it is not straightforward to find a parsimonious model that can efficiently handle the cross-correlation between them (see Section \ref{subsubsection:challenges} below for further discussion and Section \ref{subsection:RedDimApproaches} for our solution).

\subsubsection{Computational and Inferential Challenges} \label{subsubsection:challenges}

The emulation and calibration problems for both the past grounding line positions and the modern ice-no ice binary patterns face computational and inferential challenges when the length of time series $n$ and the number of spatial locations $m$ are large. For both of these problems, the likelihood evaluation in the emulation step involves Cholesky decomposition of $nq \times nq$ and $mp \times mp$ covariance matrices, which scales as $\mathcal{O}(n^3q^3)$ and $\mathcal{O}(m^3p^3)$ respectively. For our WAIS model calibration problem, this requires $\frac{1}{3}n^3q^3=1.1\times10^{17}$ and $\frac{1}{3}m^3p^3=2.6 \times 10^{18}$ flops of computation for each likelihood evaluation,  which translate to about 28,000 hours and 220,000 hours on a high-performance single core. Moreover, emulation and calibration using the modern ice-no ice patterns poses additional inferential difficulties. In particular, we need to compute $mp=1,988,750$ logits for the model output in the emulation step and $2m=6,364$ logits for the observational data. The challenge is therefore to ensure that the problem is well-posed by constraining the logits, while at the same time retaining enough flexibility in the model. The problem is even more complicated due to the dependence between the discrepancy terms $\bdelta$ and $\bphi$, because we need to estimate $n \times m = 1,501 \times 3,182$ correlation coefficients between the elements in those two terms.

%In particular, the inferential problem is ill-posed due to the fact that the sizes of data sets are $mp$ and $m$ for emulation and calibration respectively, which requires that we compute $mp=1,988,750$ logits for the model output in the emulation step and $2m=6,364$ logits for the observational data. 
 
\subsection{Reduced-dimension Approach} \label{subsection:RedDimApproaches}

In this subsection, we discuss our reduced-dimension approaches to mitigate the computational and inferential challenges described above.

\subsubsection{Dimension Reduction using Principal Components}

The first step is to reduce the dimensionality of model output via principal component analysis. For the model output matrix of the past grounding line positions $\bY_1$ we find the $J_1$-leading principal components by treating each of its columns (i.e. output for each time point) as different variables and rows (i.e. output for each parameter setting) as repeated observations. For computational convenience we rescale the principal component scores by dividing them by the square roots of their corresponding eigenvalues so that their sample variances become 1.  We denote the $j$th rescaled principal component scores at the parameter setting $\btheta$ as $Y_1^R(\btheta,j)$ and the $q \times J_1$ matrix that contains all the principal component scores for the design points $\btheta_1,\dots,\btheta_q$ as $\bY_1^R=\left\{ Y_1^R(\btheta_i,j) \right\}$ with its rows for different parameter settings and columns for different principal components. Similarly, for the model output matrix of the modern ice-no ice binary patterns patterns  $\bY_2$ we form a $p \times J_2$ matrix $\bY_2^R=\left\{ Y_2^R(\btheta_i,j) \right\}$ of $J_2$ leading logistic principal components in the same way, where $Y_2^R(\btheta,j)$ is the $j$th logistic principal component at the parameter setting $\btheta$.

We use $J_1=20$ principal components for the past grounding position output and $J_2=10$ for the modern binary pattern output. Through a cross validation experiment described below in Section \ref{section:results}, we have found that increasing the number of principal components does not improve the emulation performance. We have also confirmed that the principal component score surfaces vary smoothly in the parameter space and hence Gaussian process emulation is a suitable approach to approximating them (Figures S2-S7 in the Supplement).

We display the first three principal components for modern binary spatial pattern in Figure S8 and past grounding line position time series in Figure S9. The first three principal components for the modern binary spatial pattern show that the most variable patterns between parameter settings are (i) the overall ice coverage in the inner part of the Amundsen Embayment, which determines whether there is a total collapse of ice sheet in this area, (ii) the grounding line pattern around the edge of Amundsen Sea Embayment, (iii) and the ice coverage around the Thurston island. The first three principal components for the past grounding line position time series indicate that the most variable patterns between the input parameter settings are (i) the grounding-line retreat occurring between 15,000 and 7,000 years ago, (ii) retreat occurring until around 9,000 years ago, followed by strong re-advance until 3,000 years ago (red dashed curve), and (iii) a quasi-sinusoidal advance and retreat (black dashed-dotted curve) that spans the entire time period.

\subsubsection{Emulation using Principal Components}

 In the emulation step, our principal component-based approach allows us to circumvent expensive matrix computations in likelihood evaluation (Section \ref{subsubsection:challenges}, above) by constructing emulators for each principal component separately. For the $j$th principal component of $\bY_1$, we fit a Gaussian process model to $Y_1^R(\btheta_1,j),\dots,\allowbreak Y_1^R(\btheta_q,j)$ ($j=1,\dots,J_1$) with 0 mean and the following covariance function:

\begin{equation*} \label{eqn:exp_cov1}
\mbox{Cov}(Y_1^R(\btheta_k,j),Y_1^R(\btheta_l,j))=\kappa_{1,j}\exp\left(-\sum_{i=1}^{4}
\frac{|\theta_{ik}-\theta_{il}|}{\phi_{1,ij}} \right)+\zeta_{1,j} 1(\btheta_k=\btheta_l) ,
\end{equation*}
with $\kappa_{1,j}, \phi_{1,1j},\dots,\phi_{1,4j}, \zeta_{1,j} >0$ by finding the MLEs $\hat{\kappa}_{1,j}, \hat{\phi}_{1,1j},\dots,\hat{\phi}_{1,4j}$, and $\hat{\zeta}_{1,j}$. The computational cost for likelihood evaluation is reduced from $\mathcal{O}\left( q^3 n_1^3 \right)$ to $\mathcal{O}\left(J_1 q^3\right)$. The resulting $J_1$ Gaussian process models allow us to interpolate the values of the principal components at any new value of $\btheta$. We denote the collection of the predicted values  given by these Gaussian process models for parameter setting $\btheta$ as $\boldeta(\btheta,\bY_1^R)$. Similarly we construct Gaussian process models for the logistic principal components $Y_2^R(\btheta_1,j),\dots,Y_2^R(\btheta_p,j)$ ($j=1,\dots,J_2$) with mean 0 and the covariance function

\begin{equation*} \label{eqn:exp_cov2}
\mbox{Cov}(Y_2^R(\btheta_k,j),Y_2^R(\btheta_l,j))=\kappa_{2,j}\exp\left(-\sum_{i=1}^{4}
\frac{|\theta_{ik}-\theta_{il}|}{\phi_{2,ij}} \right)+\zeta_{2,j} 1(\btheta_k=\btheta_l) ,
\end{equation*}
with $\kappa_{2,j}, \phi_{2,1j},\dots,\phi_{2,4j}, \zeta_{2,j} >0$, by finding the MLEs $\hat{\kappa}_{2,j}, \hat{\phi}_{2,1j},\dots,\hat{\phi}_{2,4j}$, and $\hat{\zeta}_{2,j}$. This reduces the computational cost for likelihood evaluation  from $\mathcal{O}(m^3p^3)$ to $\mathcal{O}(J_2 p^3)$. Moreover, our approach requires computing only $p\times J_2$ logistic principal components and hence eliminates the need for computing $mp$ logits. As above, we let $\bpsi(\btheta,\bY_2^R)$ be the collection of the values of logistic principal components at any new value of $\btheta$ interpolated by the Gaussian process models. 

\subsubsection{Dimension-reduced Calibration} \label{subsubsectoin:RedDimCalib}

In the calibration step, we use basis representations for the observational data sets using the emulators for the principal components constructed above to mitigate the computational and inferential challenges explained in Section \ref{subsubsection:challenges}.  For the observed past grounding line positions $\bZ_1$ we set up the following linear model:

\begin{align} \label{equation:obs1}
\bZ_1=\bK_{1,y} \boldeta(\btheta,\bY_1^R) + \bK_{1,d} \bnu_1 + \bepsilon_1,
\end{align}
where $\bK_{1,y}$ is the $n \times J_1$ matrix for the eigenvectors for the leading principal components rescaled by square roots of their corresponding eigenvalues, $ \bK_{1,d} \bnu_1$ is a low-rank representation of the discrepancy term $\bdelta$, with an $n\times M$ basis matrix $\bK_{1,d}$ and its $M$-dimensional random coefficient vector $\bnu_1 \sim N(\mathbf{0}, \alpha_1^2 \mathbf{I}_{M})$ $(\alpha_1^2>0)$, and $\bepsilon_1$ is a vector of $n$ i.i.d random errors with mean 0 and variance $\sigma_{\epsilon}^2>0$ (see Section \ref{subsection:discrepancy} below for the details on specifying $\bK_{1,d}$). Inferring parameters using dimension-reduced observational data computed based on the representation in \eqref{equation:obs1} leads to a significant computational advantage by reducing the computational cost for likelihood evaluation from the order of $\mathcal{O}(n^3)$ to the order of $\mathcal{O}\left((J_1+M)^3\right)$ (see Appendix A for details).

The idea of using principal components and kernel convolution in calibration is similar to the approach described in \citet{higdon2008}. However, our approach enables a faster computation by emulating each principal component separately and formulating the calibration model in terms of the dimension-reduced observational data $\bZ_1^R$; the approach in \citet{higdon2008} retains the original data $\bZ_1$ and hence their computational gains are primarily due to more efficient matrix operations. Moreover, we use a two stage approach \citep[cf.][]{bayarri2007computer,bhat2013inferring,chang2013fast}, which separates the emulation and the calibration steps, to reduce the identifiability issues between the parameters in the emulator and the discrepancy term.

For the modern observed ice-no ice binary patterns $\bZ_2$ we set up the following linear model for the logits:
\begin{equation} \label{equation:obs2}
\blambda=\bK_{2,y} \bpsi(\btheta,\bY_2^R)+  \bK_{2,d} \bnu_2,
\end{equation}
where $\bK_{2,y}$ is the $m\times J_2$ eigenvectors for the leading logistic principal components and $\bK_{2,d}$ is an $m \times L$ basis matrix with $L$-dimensional random coefficients $\bnu_2 \sim N(\mathbf{0}, \alpha_2^2 \mathbf{I}_L)$ (see Section \ref{subsection:discrepancy} below for the details on specifying $\bK_{2,d}$).  This basis representation also reduces the cost for matrix computation from $\mathcal{O}(m^3)$ to $\mathcal{O}(J_2 p^3)$. More importantly, using this basis representation reduces the number of logits that need to be estimated from $2m$ to $J_2+L$, and hence makes the calibration problem well-posed. Using the model in \eqref{equation:obs1} and \eqref{equation:obs2} and additional prior specification we can set up the posterior density and estimate the input parameters $\btheta$ while accounting for the uncertainty in other parameters via MCMC using the standard Metropolis-Hastings algorithm. We describe posterior density specification in more details in Appendix B.

We need to consider the dependence between $\bZ_1$ and $\bZ_2$ to use the information from both data sets simultaneously. As discussed above we model the dependence between the emulators through the input parameter $\btheta$. We also capture the dependence between the discrepancy terms through the $M \times L$ cross correlation matrix $\mathbf{R}_\nu$ between $\bnu_1$ and $\bnu_2$, where the $(i,j)$th element of $\mathbf{R}_\nu$ is $\rho_{\nu,ij}=\mbox{Cov}(\nu_{1i},\nu_{2j})$ and $\nu_{1i}$ and $\nu_{2j}$ are respectively the $i$th and $j$th elements of $\bnu_1$ and $\bnu_2$ (See Appendix A and B for further details). This greatly reduces the inferential issue by reducing the number of cross-correlation coefficients that need to be estimated from $m n$ to $M L$.

%Moreover, using reduced-dimension approach also mitigates the inferential challenge for calibration using binary patterns explained above  by imposing a low dimensional structure for the logits of model output and observational data that keeps the number of parameters to be estimated reasonably low while allowing for enough flexibility to accurately represent the binary patterns from model output and observational data.

\subsection{Model-Observation Discrepancy} \label{subsection:discrepancy}

For a successful computer model calibration it is important to find good discrepancy basis matrices $\bK_{1,d}$ and $\bK_{2,d}$  that allow enough flexibility in representing the model-observation discrepancies while avoiding identifiability issues in estimating $\btheta$ \citep[cf.][]{brynjarsdottir2014learning}. To define the discrepancy basis matrix for the past grounding line positions $\bK_{1,d}$ we use a kernel convolution approach using $M<n$ knot points $a_1,\dots,a_M$ that are evenly distributed between $t_1$ and $t_n$. We use the following exponential kernel function to define the correlation between $t_1,\dots,t_n$ and $a_1,\dots,a_M$:
\begin{equation} \label{eqn:kerneldisc}
\left\{\bK_{1,d}\right\}_{ij}=\exp\left(-\frac{\left|t_i-a_j\right|}{\phi_{1,d}}\right),
\end{equation}
with a fixed value $\phi_{1,d}>0$. The basis representation based on this kernel function enables us to represent the general trend in model-observation discrepancy using a small number of random variables for the knot locations. Note that the discrepancy term constructed by kernel convolution can confound the effect from model parameters and thus cause identifiability issues; any trend produced by  $\bK_{1,y} \boldeta(\btheta,\bY_1^R)$ can be easily approximated by $\bK_{1,d} \bnu_1$ and therefore one cannot distinguish the effects from these two terms \citep{chang2013fast}. To avoid this issue, we replace $\bK_{1,d}$ with its leading eigenvectors, which corresponds to regularization given by ridge regression \citep[see][pg. 66]{trevor2009elements}. In the calibration experiment in Section \ref{section:results} we chose the value of $\phi_{1,d}$ as 750 (years), the number of knots as $M=1,500$, and the the number of eigenvectors as 300 and confirmed that using the discrepancy term based on these values leads to a reasonable calibration result by a simulated example (Section \ref{subsection:SimulatedExample}). We have also found that different settings for these values lead to similar calibration results and hence inference for $\btheta$ is robust to choice of these values \cite[cf.][]{chang2013fast}.

The identifiability issue explained above is further complicated for the modern ice-no ice binary patterns because binary patterns provide even less information for separating the effects from the input parameters and the discrepancy term than continuous ones do. Through some preliminary experiments (not shown here) we found that the regularization introduced above does not solve the identifiability issue for binary patterns. Therefore we use an alternative approach to construct the discrepancy basis $\bK_{2,d} $, which is based on comparison between model runs and observational data \citep{chang2015binary}. In this approach $\bK_{2,d}$ has only one column (i.e. $L=1$) and therefore the matrix is reduced to a column vector $\bk_{2,d}$ and its coefficient vector $\bnu_2$ becomes a scalar $\nu_2$. For the $j$th location $\bs_j$ we calculate the following signed mismatch between the model and observed binary outcomes:

\begin{equation*}
r_j=\frac{1}{p}\sum_{i=1}^p  \mbox{sgn}(Y_2(\btheta_i,\bs_j)-Z_2(\bs_j)) I(Y_2(\btheta_i,\bs_j)\ne Z_2(\bs_j)),
\end{equation*}
where $\mbox{sgn}(.)$ is the sign function. If $|r_j|$ is greater than or equal to a threshold value $c$, we identify $\bs_j$ as a location with notable discrepancy and define the corresponding $j$th element of $\bk_{2,d}$ as  the logistic transformed $r_j$, $\log(\frac{1+r_j}{1-r_j})$. If $|r_j|<c$, we assume that the location $\bs_j$ shows no notable model-observation discrepancy and set the $j$th element of $\bk_{2,d}$ as 0. Choosing a too large value of $c$ results in inaccurate discrepancy representation by ignoring important patterns in the model-observation discrepancy while a too small value of $c$ causes identifiability issues between the input parameters $\btheta$ and the discrepancy term. Based on experiments with different model runs and observational data sets \cite[cf.][]{chang2015binary} we found that setting $c$ to be 0.5 gives us a good balance between accurate discrepancy representation and parameter identifiability.

\section{Implementation Details and Results} \label{section:results}
We calibrate the PSU3D-ICE model (Section \ref{section:ModelAndObs}) using our reduced dimension approach (Section \ref{subsection:RedDimApproaches}). We first verify the performance of our approach using simulated observational data sets (Section \ref{subsection:SimulatedExample}) and then move on to actual calibration using real observational  data sets to estimate the input parameters and generate WAIS volume change projections (Section \ref{subsection:RealObservations}).

\begin{figure}
\centering
\includegraphics[scale=0.32]{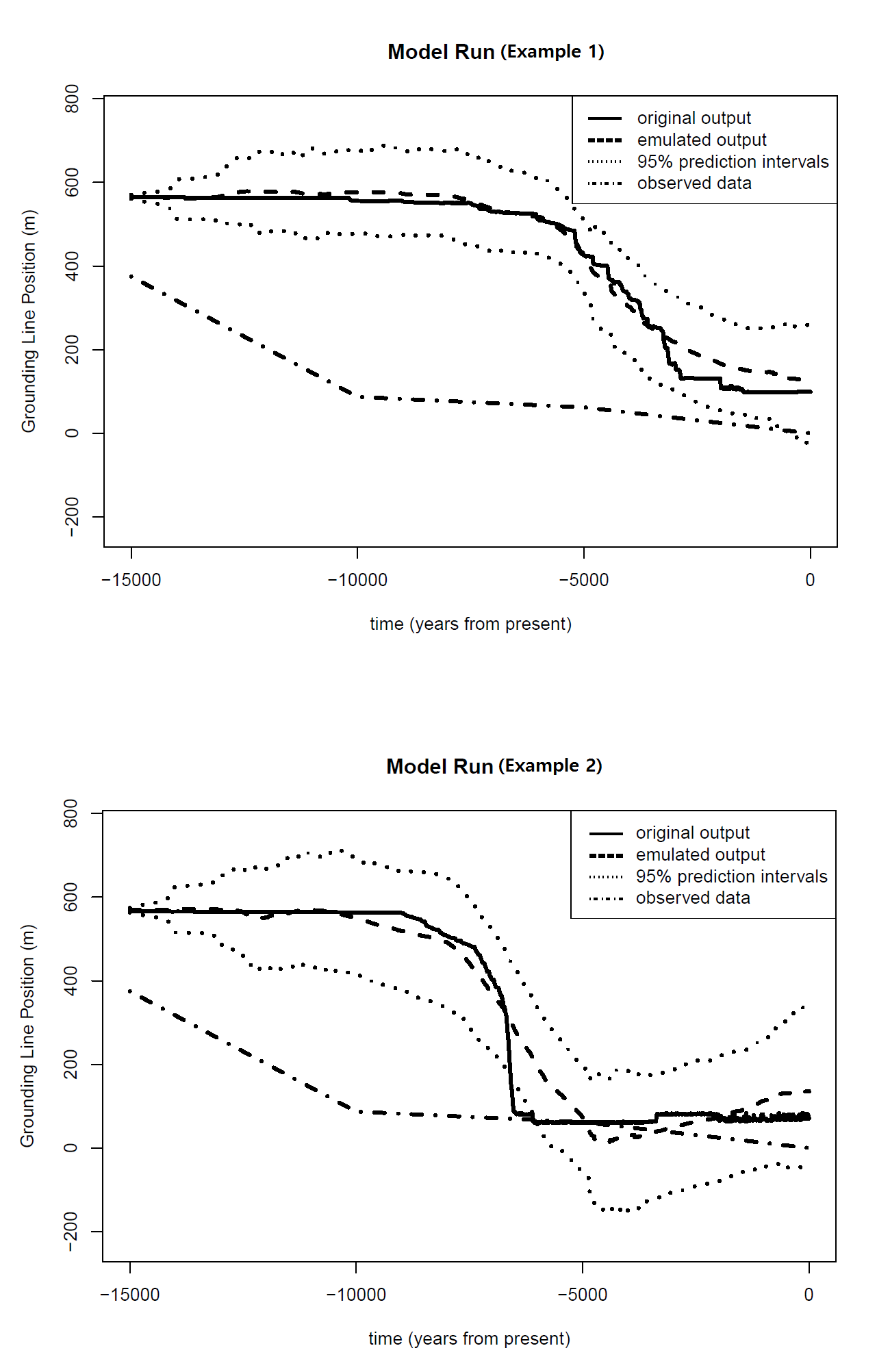}
\caption{Results for two example parameter settings from the leave-out experiment to verify the performance of the emulator for past grounding line positions. Results for other parameter settings are qualitatively similar to the results shown here.  In general, the emulated grounding positions are similar to those from the actual model runs. For comparison we have also added the reconstructed grounding line position observations.}
\label{fig:GroundingEmulation}
\end{figure}

\begin{figure}
\centering
\includegraphics[scale=0.29]{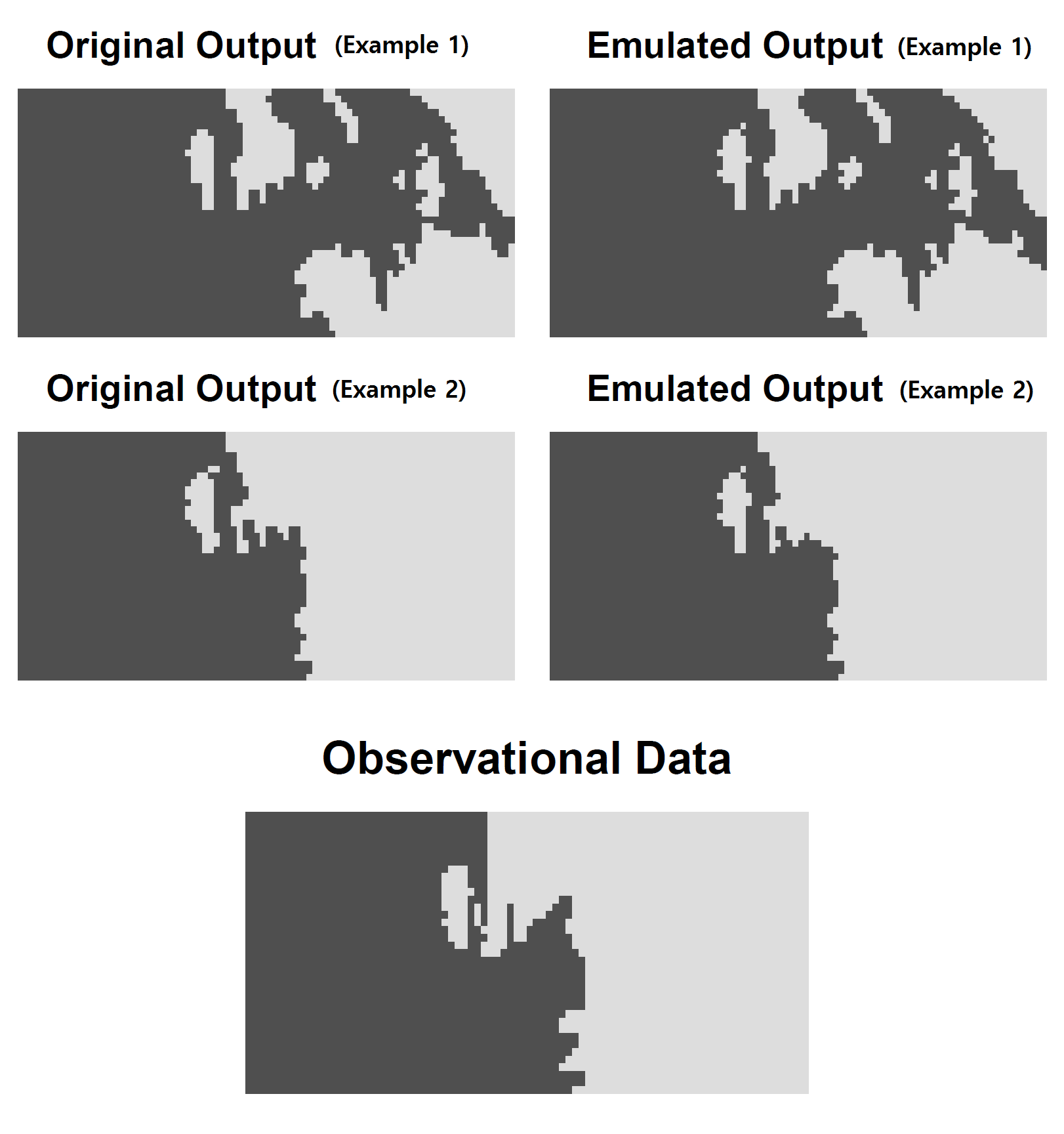}
\caption{Results for two example parameter settings from the leave-out experiment to verify the performance of the emulator for modern binary patterns (light gray for grounded ice and dark gray for no grounded ice). Results for other parameter settings are similar to the ones presented here. In general, the emulator can accurately approximate the binary patterns from the actual model runs. For comparison we have also included the observed modern binary patten.}
\label{fig:BinaryEmulation}
\end{figure}

Before calibration, we verified the performance of our emulators through separate leave-out experiments for each emulator $\boldeta$ and $\bpsi$. In each experiment we leave out a group of model runs around the center of the parameter space from the ensemble and try to recover them using an emulator constructed based on the remaining model runs. We have left out 82 model runs for the modern binary ice patterns and 60 for the past grounding line positions since we use a smaller  number of model runs to emulate  the grounding line position output ($q=461$) than the modern binary pattern output ($p=625$). 
%via leave-10-out cross-validation. For each step of the cross-validation we randomly choose 10 different model runs and their corresponding parameter settings. We then build emulators based on the remaining model runs and interpolate the model outputs at the parameter settings that were left out using the emulator. We repeat this step to make the same comparison for all the parameter settings in the ensemble and compare the emulated outputs and the actual model outputs throughout all the parameter settings. 
Some examples of the comparison results are shown in Figures \ref{fig:GroundingEmulation} (for the past grounding line positions) and \ref{fig:BinaryEmulation} (for the modern binary patterns). The results show that our emulators can  approximate the true model output reasonably well.

\subsection{Simulated Example} \label{subsection:SimulatedExample}
In this subsection we describe our calibration results based on simulated observations to study how accurately our method recovers the assumed true parameter setting and its corresponding true projected ice volume change. The assumed-true parameter setting that we choose to use here is OCFAC=0.5, CALV=0.5, CRH=0.5, and TAU=0.4 (rescaled to [0,1]), which correspond to OCFAC=1 (non-dimensional), CALV=1 (non-dimensional), CRH=$10^{-7}$ (m/year Pa$^2$), and approximately TAU=2.6 (k year) in the original scale. This is one of the design points that is closest to the center of the parameter space.

\begin{figure}
\centering
\includegraphics[scale=0.6]{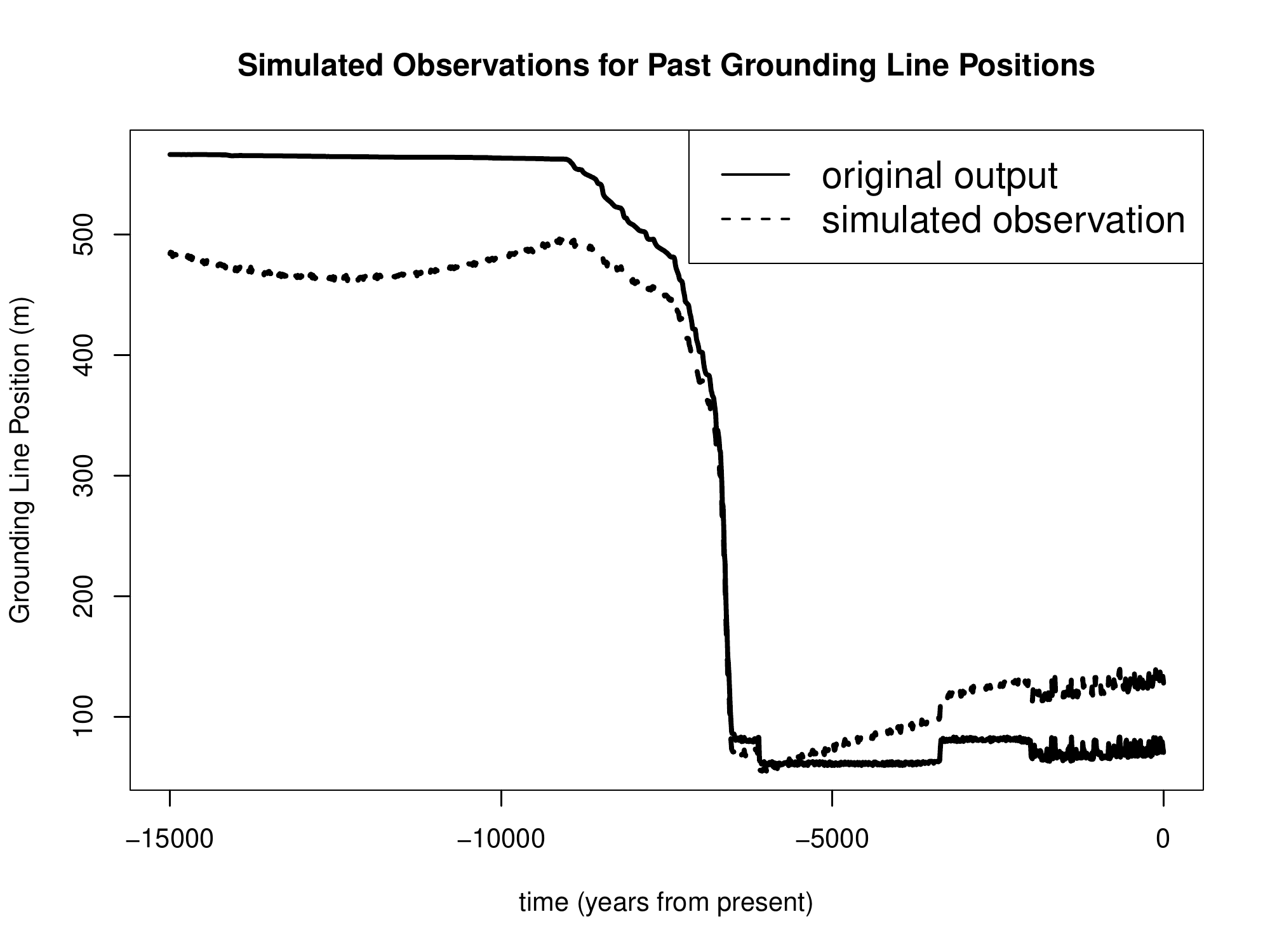}
\includegraphics[scale=0.4]{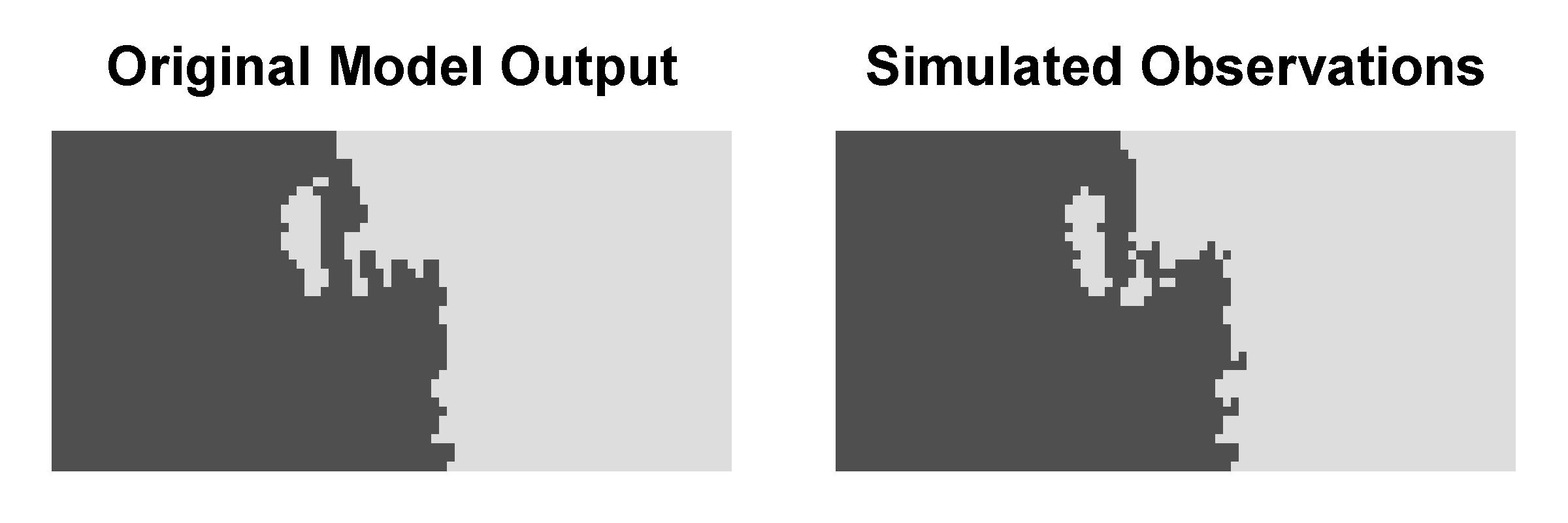}
\caption{The original model outputs at the assumed true parameter setting and the simulated observations used for the simulated example in Section \ref{subsection:SimulatedExample}. The upper plot is for the past grounding line positions and the lower plots are for the modern binary patterns (light gray for grounded ice and dark gray for no grounded ice).}
\label{fig:OriginalVsSimulatedObs}
\end{figure}

To represent the presence of model-observation discrepancy  we contaminate the model outputs at the true parameter setting with simulated structural errors. We generate simulated errors for the past grounding line positions from a Gaussian process model with zero mean and the covariance defined by the squared exponential function with a sill of 90 m and a range of 10,500 years. The generated errors represent a situation where the model-observation discrepancy varies slowly and has a persistent trend over time. We generate the discrepancy for the modern binary patterns in a manner that makes them realistic. Our approach is therefore as follows. We use the original model output and observational data for the ice thickness pattern used to derived the modern binary patterns. We first choose the ``best'' 90\% model runs (i.e. exclude the ``worst'' 10\% model runs) that are closest to the modern observations in terms of the mean squared differences and take pixel-wise averages for the selected model runs to derive a common thickness pattern. We then subtract the common thickness pattern from the observational data to derive the common discrepancy pattern in terms of the ice thickness. By subtracting this discrepancy pattern from the ice thickness pattern for the assumed truth and dichotomizing the resulting thickness pattern into a binary ice-no ice pattern, we obtain simulated observations for the modern binary pattern. We illustrate the resulting simulated observations in Figure \ref{fig:OriginalVsSimulatedObs}. 

Depending on the computational environment, the emulation step typically takes several minutes to half an hour. The computation in the calibration step requires about 240 hours to obtain an MCMC sample of size 70,000. The values of MCMC standard errors \citep{jones2006fixed,flegal2008markov} suggest that the sample size is large enough to estimate the posterior mean. We have also compared the marginal densities estimated from the first half the MCMC chain and the whole chain and confirmed that the chain is stable.

\begin{figure}
\centering
\includegraphics[scale=0.7]{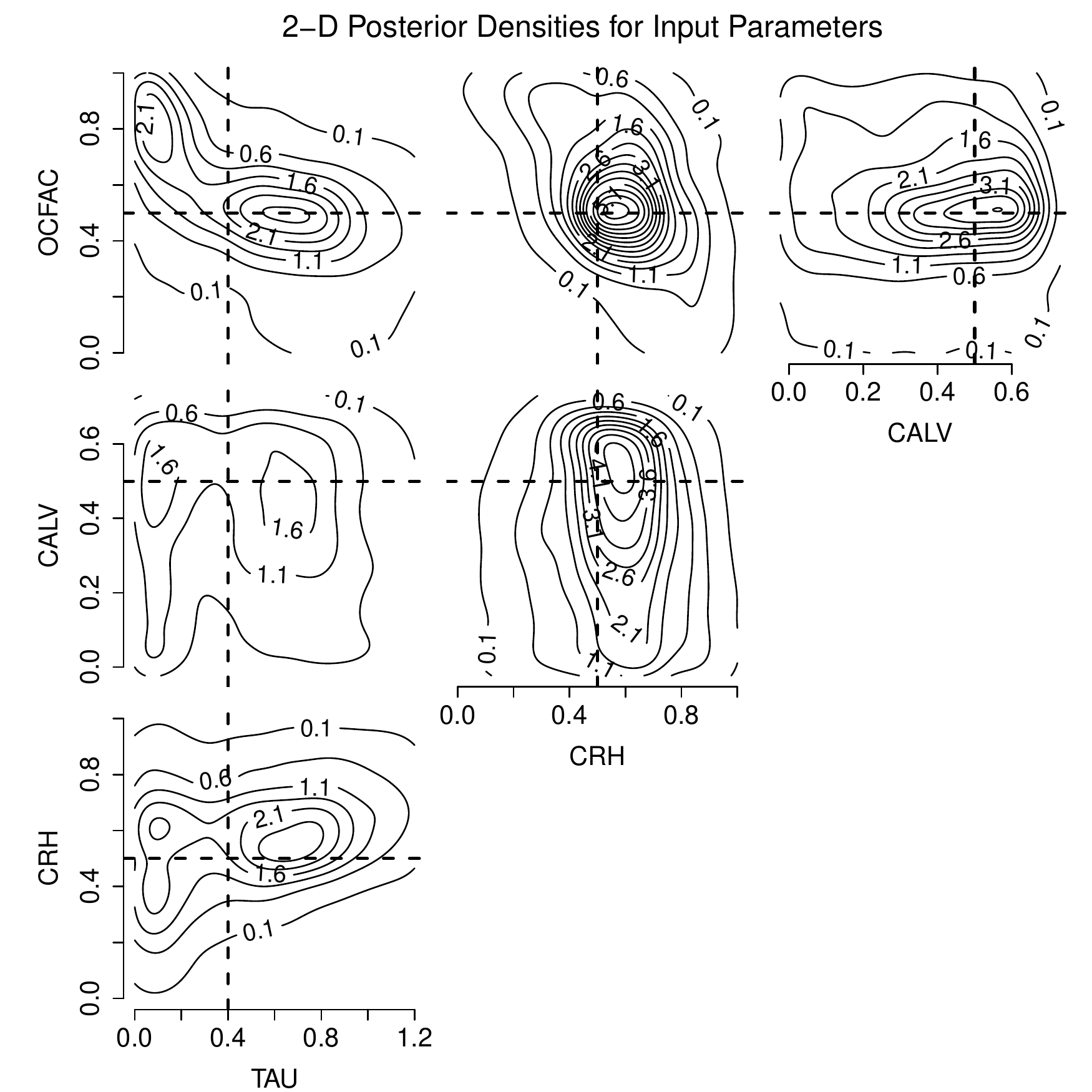}
\caption{Posterior density for input parameters from the perfect model experiment described in Section \ref{subsection:SimulatedExample}. The black dashed lines represent the assumed parameter settings. The result indicates that our method can recover the true input parameter setting by assigning high posterior density around the truth. The joint density in the upper left panel shows that OCFAC and TAU have a strong non-linear interaction.}
\label{fig:SimulatedDensity}
\end{figure}

The probability density plots in Figure \ref{fig:SimulatedDensity} show that the posterior has  a high probability mass around the assumed truth and hence indicate that our approach can recover the true parameter setting. However, the results also show that parameter settings other than the assumed truth have high probability densities, suggesting that settings other than the assumed truth can yield  similar outputs to the simulated observations. Interestingly the joint posterior density for TAU and OCFAC shows that these two parameters have a strong nonlinear interaction with each other.  

\begin{figure}
\centering
\includegraphics[scale=0.5]{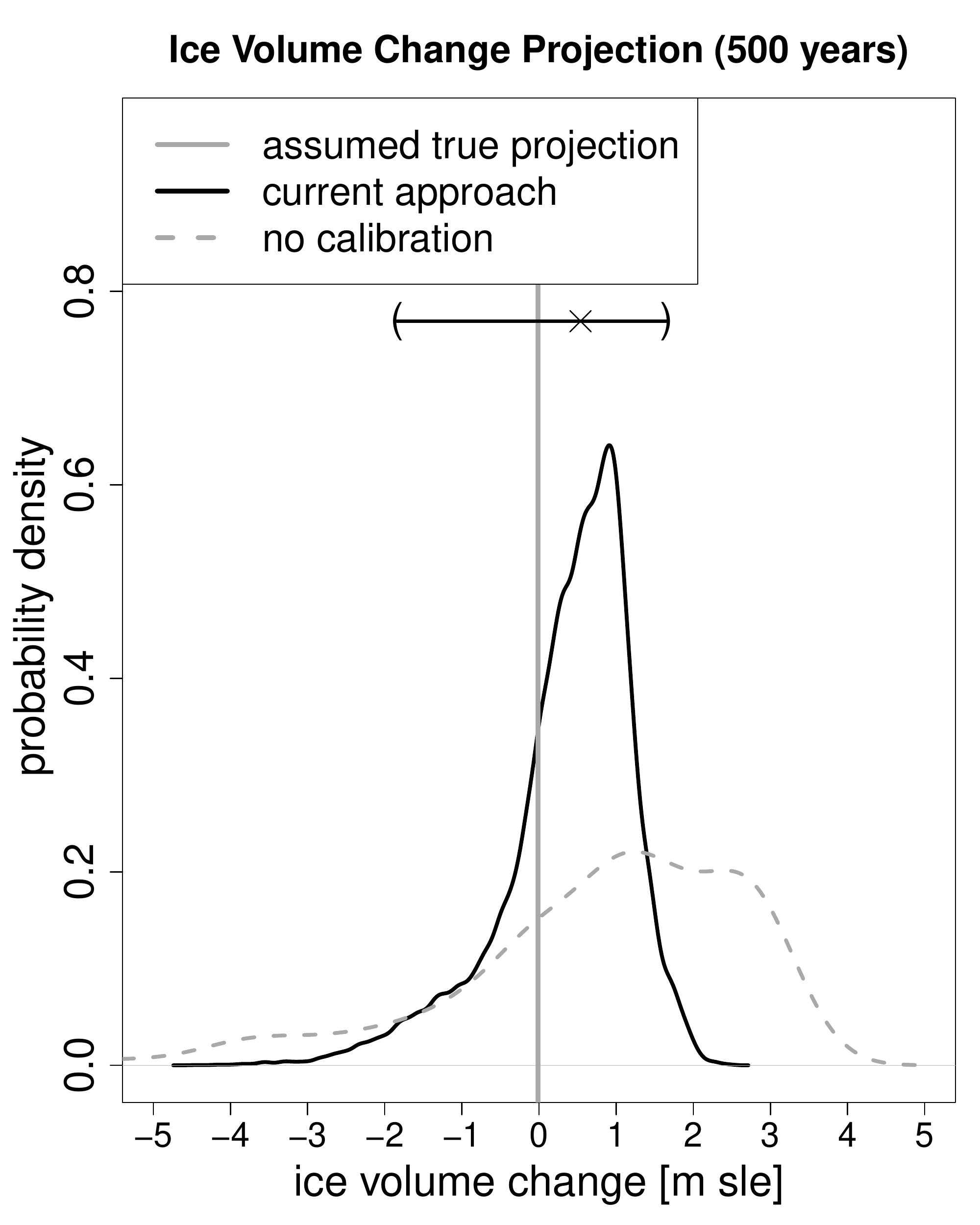}
\caption{Predictive distribution for the ice volume change projection based on the simulated observations described in Section \ref{subsection:SimulatedExample}. The gray solid line shows the assumed truth. The black solid and the gray dashed lines respectively show the predictive distributions with and without calibration using our approach. The result shows that our approach can recover the assumed truth and significantly reduce the uncertainty in projection.}
\label{fig:SimulatedProjection}
\end{figure}

Figure \ref{fig:SimulatedProjection} shows the resulting projection for ice volume change for 500 years from present. To generate the projections we first built an emulator to interpolate the ice volume change values at the existing 625  parameter settings and then converted our MCMC chain for $\btheta$ into a Monte Carlo sample for the ice volume change values using the emulator. The result also confirms that our method accurately recovers the assumed-true projection with reasonable uncertainty.

\subsection{Calibration Using Real Observations} \label{subsection:RealObservations}

We are now ready to calibrate the PSU3D-ICE model using the real observational datasets described in Section \ref{section:ModelAndObs}.  Our main goal is to investigate whether using information on past grounding line position in our calibration leads to reduced uncertainty and better-constrained projections. To this end, we calibrate the PSU3D-ICE model based on (i) only the modern binary patterns and (ii)  both information sources. For (i) we conduct calibration using only the part of the posterior density that is related to $\bY_2$ and $\bZ_2$ and for (ii) we use the entire posterior density. We have obtained MCMC samples with 100,000 for (i) and 47,000 iterations for (ii) and checked its convergence and standard errors, as above. 

\begin{figure}
\centering
\begin{subfigure}{0.7\textwidth}
        \includegraphics[scale=0.45]{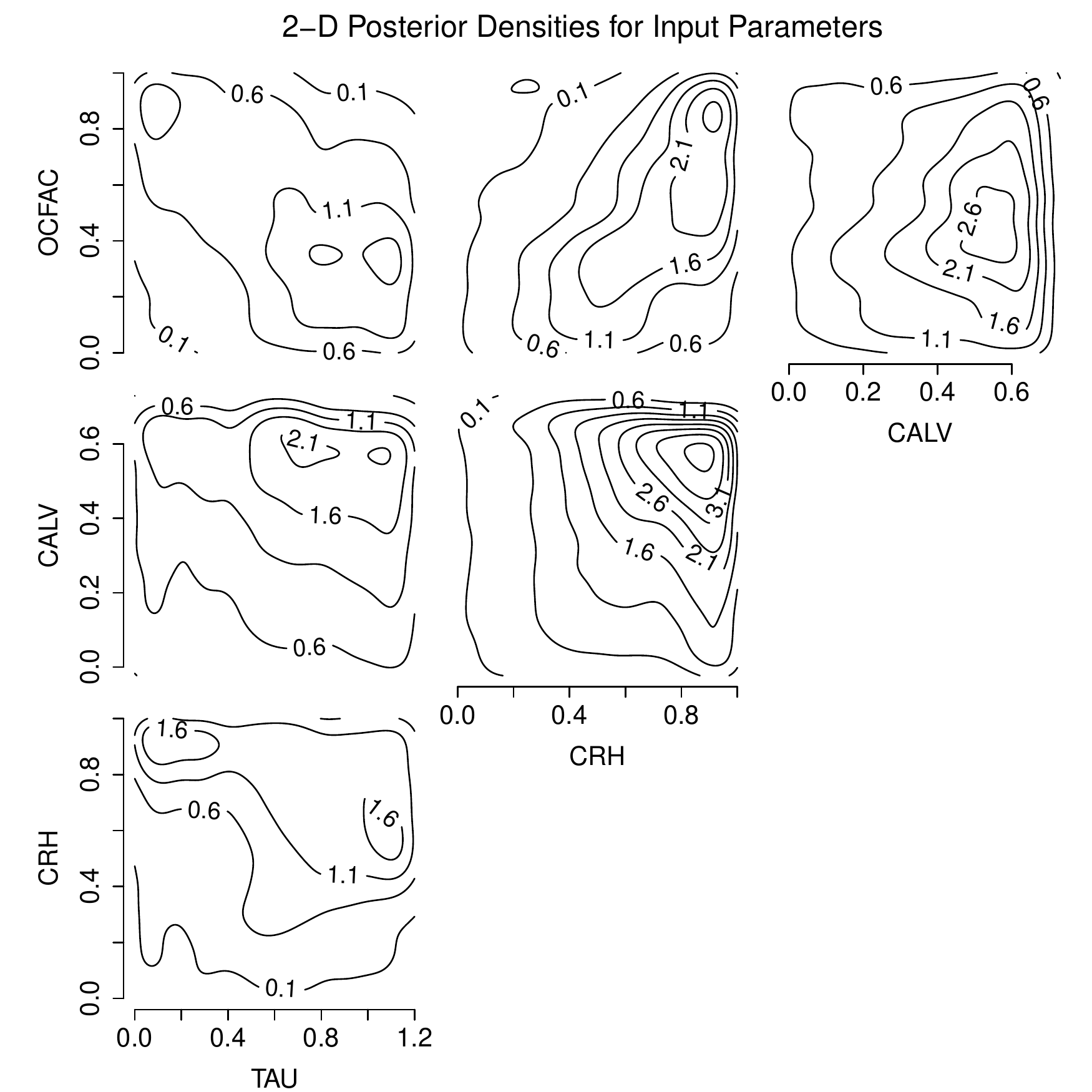}
\caption{Modern Binary Patterns Only}
\end{subfigure}
\begin{subfigure}{0.7\textwidth}
                \includegraphics[scale=0.45]{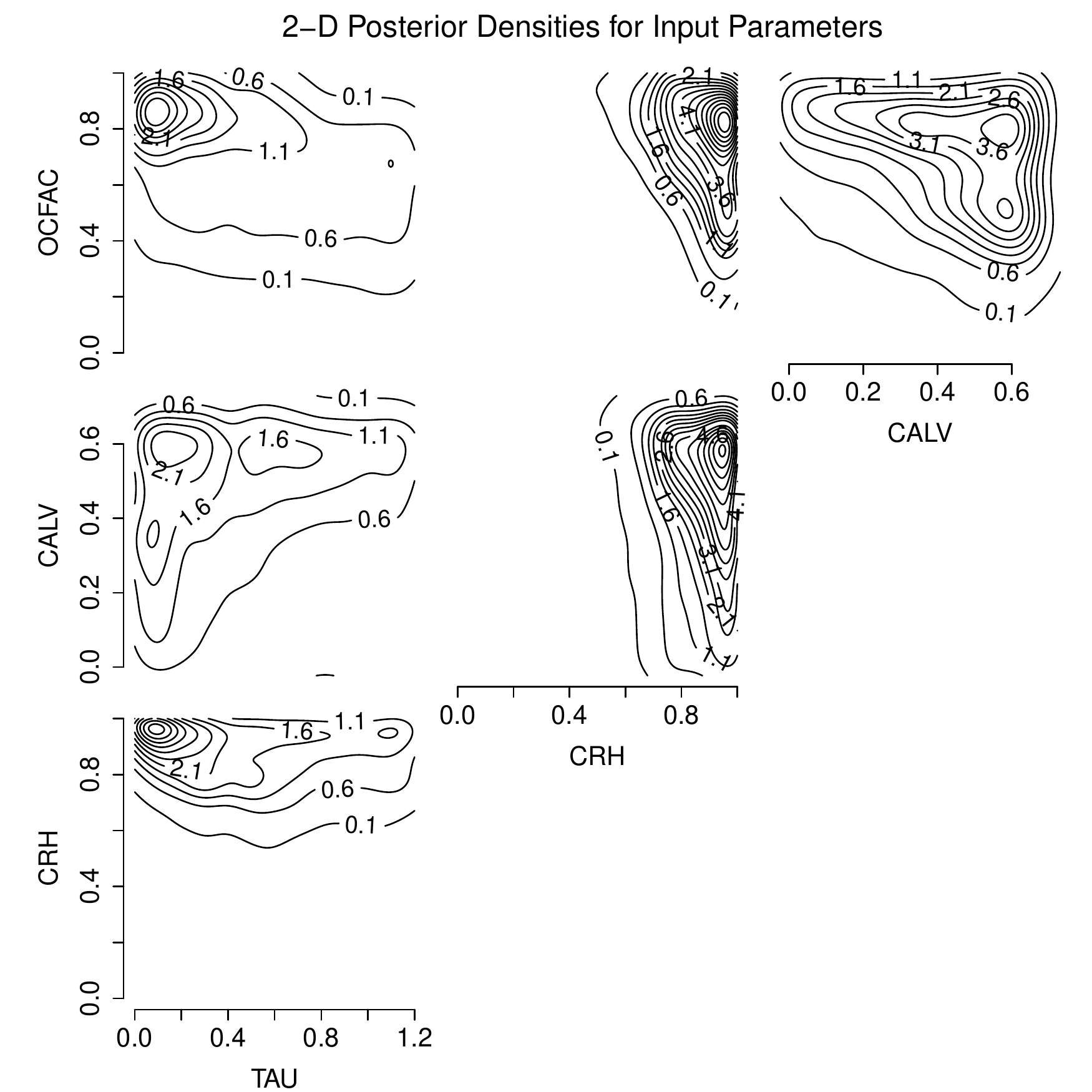}
                \caption{Modern Binary Patterns and Past Grounding Line Positions}
\end{subfigure}
\caption{Posterior density for input parameters based on the actual observational data sets (see Section for details Section \ref{subsection:RealObservations}). Using both information sources leads to significantly less uncertainty in estimating input parameters.}
\label{fig:ObsDensity}
\end{figure}

\begin{figure}
\centering
\includegraphics[scale=0.5]{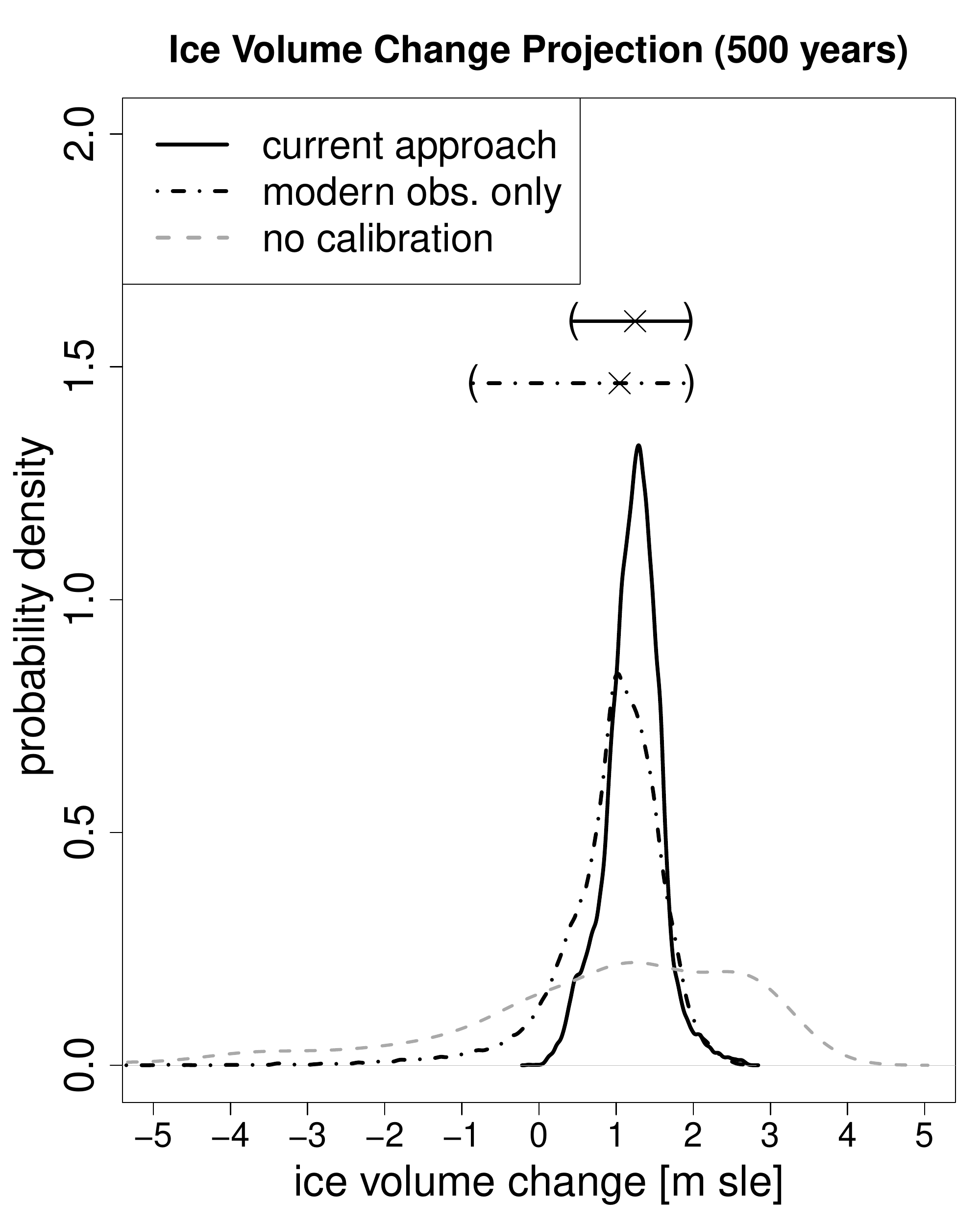}
\caption{Predictive distribution for the ice volume change projection based on the real observations described in Section \ref{subsection:RealObservations}. The black solid line shows the predictive distribution based on our approach using both the past grounding line positions the modern binary patterns and the black dashed and dotted line represents the result based on only the modern binary patterns. The gray dashed line shows the projection without calibration. The results show that using the past grounding line leads to a significantly sharper projection by removing the unrealistic ice volume increase in the results solely based on the modern observations.}
\label{fig:ObsProjection}
\end{figure}

The results clearly show the utility of using past grounding line positions in calibrating the PSD3D-ICE model. Figure \ref{fig:ObsDensity} shows that using the past grounding line positions in our calibration drastically reduces parametric uncertainty. The ice volume change projection based on both information sources also has significantly less uncertainty than that based only on the modern binary patterns (Figure \ref{fig:ObsProjection}). In particular, using the grounding line positions in calibration eliminates the probability mass below 0 m sea level equivalent in the predictive distribution. This improvement is due to the fact that some  parameter settings result in very unrealistic ice sheet behavior from the last glacial maximum to the present day, but give modern ice coverage patterns that are  close to the current state. See the next subsection for further discussion.
%To be more specific, some of the model runs used very unrealistic model parameter values, that gave extremely excessive ice-sheet retreat from last glacial maximum extents during the last 15,000 years. In these runs, grounding-line extents retreated past modern positions much too early, several thousand years ago, and receded into what is now the Antarctic interior. Due to the lagged bedrock rebound feedback, grounding-line depths then shallowed, allowing ice re-advance so that at the present day, grounding lines attained modern-like positions, thus leads to a high posterior density when compared only to modern observations. However, grounding lines in these runs continued to advance in future years, giving greater ice extents than modern and thus negative sea-level rise into the future. When additional information is used comparing with past grounding line positions, the unrealistic behavior from -10,000 to 5000 years in these runs yielded low posterior densities, essentially removing negative future sea levels from the credible interval for the projection.

Note that the calibration results in Figure \ref{fig:ObsDensity} (a) are somewhat different from the results based on the same observational data set shown by \cite{chang2015binary}, because there are 6 other parameters that are varied in the ensemble used by \cite{chang2015binary}, which are fixed by expert judgment in this experiment.

\subsection{Insights}

\begin{figure}
\centering
\includegraphics[scale=0.7]{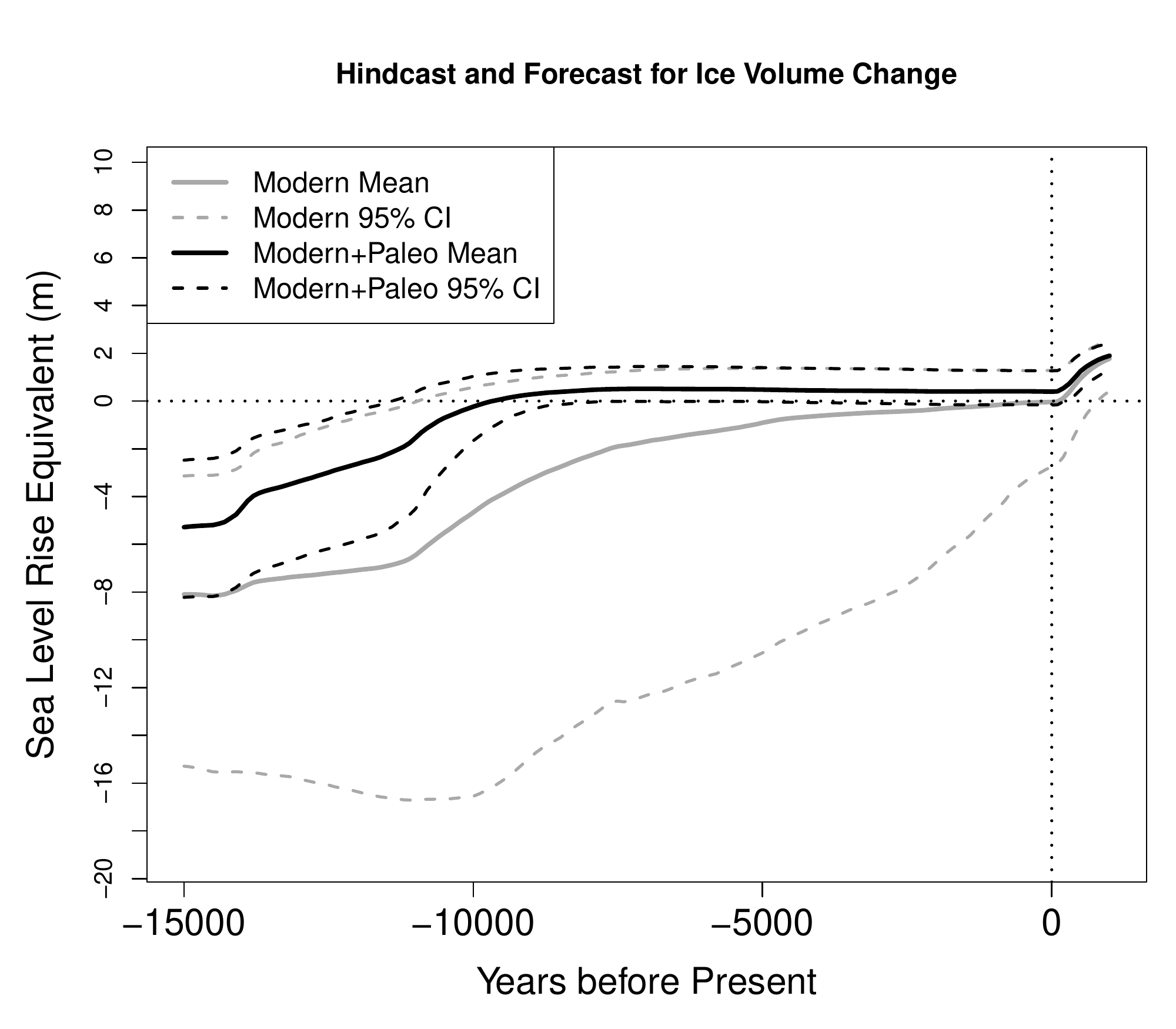}
\caption{The mean (solid lines) and point-wise 95\% predictoin limits (dashed lines) for projected ice volume changes based on the modern binary patterns only (gray) and the past grounding line positions and modern binary patterns (black). Negative values on y-axis indicate the ice volume is larger than the modern value. The prediction limits based only on modern binary patterns contain trajectories that start from excessive amount of ice volume and show very fast ice volume decay. The prediction limits based on both sources of information rule out such unrealistic trajectories.}
\label{figure:YearVsChange}
\end{figure}

The results presented in the previous subsection clearly indicate that using the past grounding line position leads to better calibration results with less parametric uncertainties and in turn sharper future ice volume change projections. The hindcast and forecast of ice volume changes based on different data sources in Figure \ref{figure:YearVsChange} clearly show the reason for this improvement. The 95\% prediction intervals show that using the information from past grounding line positions significantly reduces the uncertainties in ice volume change trajectories by ruling out the parameter settings that generate unrealistic past ice sheet behavior. In particular, some parameter settings produce ice sheets that start from a very high ice volume around 15,000 years ago and then show unrealistically rapid ice volume loss, thereby resulting in a modern ice volume that is close to the observed value. These parameter settings are the cause of the left tail of volume change distributions based on the modern binary pattern only. Using paleo data rules out these parameter settings, thereby cutting off the left tail and reducing parametric uncertainty.

\begin{figure}
\centering
\includegraphics[scale=0.7]{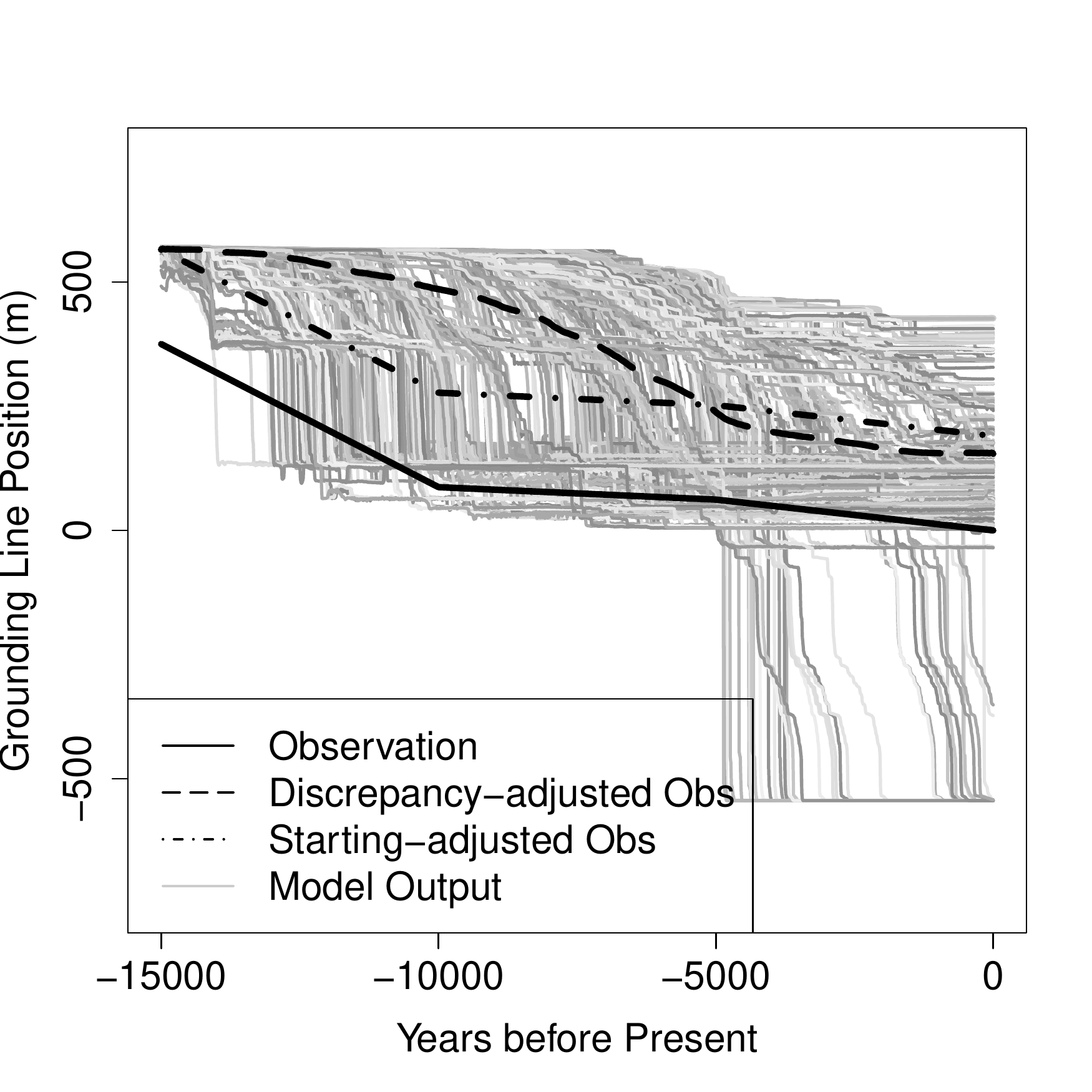}
\caption{The grounding line position time series from the observational data set (black solid line), observational data adjusted by the discrepancy term (black dashed line), observational data shifted to have the same starting point as the fully adjusted data (dashed and dotted black line), and the model outputs at the input parameter settings $\btheta_1,\dots,\btheta_p$ (gray lines). The discrepancy term shifts the observational data to match the starting grounding line position to the model outputs, which is similar to using anomalies.}
\label{figure:GroundingLinePositions}
\end{figure}

The credibility of this result of course depends on (i) how the model can reliably reproduce the past grounding line time series and (ii) how our calibration method accounts for the model-observation discrepancy. Figure \ref{figure:GroundingLinePositions} shows the grounding line position time series from the model outputs at the parameter settings $\btheta_1,\dots,\btheta_q$ and the observational data. It also shows the observational time series corrected by the discrepancy term $\bK_{1,d} \bnu_1$. The discrepancy automatically accounts for the fact that all the model time series start further inland than the observed starting points.  Hence, this is analogous to working with anomalies (difference between an observed values and a standard reference value). The use of anomalies is common in paleoclimate modeling. That is, the modeled \emph{change} in a quantity is considered more reliable than the   modeled \emph{absolute} value, because the model errors that remain constant through time cancel when differences are taken. The model change can either be the difference from the initial model state, or the difference from a ``control'' simulation of an observed state. The difference is added to the observed quantity to yield a more robust model projection. This is equivalent here to uniformly shifting the observed grounding-line positions to coincide with the mean model initial position at 15,000 years before present (Figure \ref{figure:GroundingLinePositions}).

In addition, we also show another corrected observed time series in which only the starting position is matched with the discrepancy-adjusted observed time series discussed above. The difference between this time series (dashed-dotted line in Figure \ref{figure:GroundingLinePositions}) and the fully corrected time series (dashed line in Figure \ref{figure:GroundingLinePositions}) can be viewed as the estimated discrepancy in anomaly. Although the modeled and observed anomalies clearly show a certain degree of discrepancy,  the observed anomaly still allows us to rule out the model runs that showed too small or too large total grounding line position changes over 15,000 years. Therefore the observed grounding line positions still provide useful information for reducing parametric uncertainty.

\section{Discussion and Caveats} \label{section:discussion}

\subsection{Discussion}

In this work we have proposed a computationally efficient approach to calibrating WAIS models using two different sources of information, the past grounding line positions and the modern binary spatial patters. Using the proposed approach, we have successfully calibrated the PSU-3D ice model and generated the WAIS volume change projections. Results from a simulated example indicate that our method recovers the true parameters with reasonably small uncertainties as well as provides useful information on  interactions between parameters. Results based on the real observations indicate that using the paleo-record significantly reduces parametric uncertainty and leads to better-constrained projections by removing the probability for unrealistic ice volume increase in the predictive distribution.

{\color{black}
Several recent modeling studies of Antarctic Ice Sheet variations have used heuristic approaches to study parametric uncertainties, mostly applied to ice-sheet retreat since the last glacial maximum about 15,000 years ago \citep{whitehouse2012deglacial,whitehouse2012new,briggs2013glacial,briggs2013evaluate,briggs2014data,golledge2014antarctic,maris2015model}.
Using highly aggregated data and less statistically formal frameworks,  these studies try to reduce the parametric uncertainties based on geologic data around Antarctica. By and large our results are consistent with the parameter values found in these studies, while also reducing uncertainties about the parameters. Other recent modeling studies have projected future Antarctic Ice Sheet response to anthropogenic warming in coming centuries to millennia \citep{cornford2015century,feldmann2015collapse,golledge2015multi,gomez2015sea,ritz2015potential,winkelmann2015combustion}. These models have been calibrated only over observed small-scale variations of the last few decades. Only a few of these studies use large ensembles, and none use the advanced statistical methods we have developed here, which allow for analyses based on large ensembles and unaggregated data sets. Furthermore, we are able to obtain projections as well as parameter inference in the form of genuine probability distributions, and we take into account potential data-model discrepancies that are ignored by other studies. This allows us to provide uncertainties about our estimates and projections.
}

\subsection{Caveats and Future Directions}

One caveat in our calibration model specification is that we do not take into account the dependence between the past grounding line positions and the modern binary patterns. However, we note that the past grounding line  positions and the modern binary patterns contain quite different information since two model runs with very different trajectories of past grounding line position often end up with very similar modern binary patterns; this is corroborated by an examination of cross-correlations. Developing a calibration approach based on the generalized principal component analysis that reduces the dimensionality of Gaussian and binary data simultaneously and computes common principal component scores for both data sets is one possible future direction. 

Our results are also subject to usual caveats in ice sheet modeling. For example, we use simplified atmospheric conditions for projections, which assumes that atmospheric and oceanic temperatures linearly increase until 150 years after present and stay constant thereafter. Using more detailed warming scenarios is a subject for further work. Another caveat for the present study is the use of coarse-grid global
ocean model results to parameterize past basal melting under floating
ice shelves. Fine-grid modeling of ocean circulation in Antarctic embayments
is challenging, and a topic for further work \citep[e.g.][]{hellmer2012twenty}
Another improvement will be the use of finer-scale models with higher-order
ice dynamics, which as discussed in the introduction, are not quite feasible
for the large space and time scales of this study, but should gradually become
practical in the near future.

\subsection*{Acknowledgments}
We are grateful to A. Landgraf for distributing his code for logistic PCA freely on the Web (\url{https://github.com/andland/SparseLogisticPCA}).

This work was partially supported by National Science Foundation through (1) NSF-DMS-1418090 and (2) NSF/OCE/FESD 1202632 and NSF/OPP/ ANT 1341394, (3) Network for Sustainable Climate Risk Management under NSF cooperative agreement GEO1240507, and (4) NSF Statistical Methods in the Atmospheric Sciences Network (Award Nos. 1106862, 1106974, and 1107046). WC was partially supported by (3) and (4), and MH and PA were partially supported by (1) and (3), and DP is partially supported by (1), (2) and (3).  All views, errors, and opinions are solely those of the authors.

\section*{Appendix A. Computation in Reduced-Dimensional Space}

For faster computation we infer $\btheta$ and other parameters in the model based on the following  dimension-reduced version of the observational data for the past grounding line positions:

\begin{equation*}
\bZ_1^{R}=(\bK_1^T \bK_1)^{-1} \bK_1^T \bZ_1=\left( \begin{array}{c}
\boldeta(\btheta,\bY_1^R) \\
\bnu_1
\end{array}\right) + (\bK_1^T \bK_1)^{-1} \bK_1^T \bepsilon_1,
\end{equation*}
where $\bK_1=(\bK_{1,y}~\bK_{1,d})$, which leads to the probability model 
\begin{equation} \label{eqn:Z_R prob}
\bZ_1^R | \bnu_2 \sim N \left( \left( \begin{array}{c}
\boldsymbol{\mu}_{\boldeta} \\
\boldsymbol{\mu}_{\bnu_1|\bnu_2}
\end{array}\right) ,
\left( \begin{array}{cc}
\bSigma_{\boldeta} & \mathbf{0} \\
\mathbf{0} & \bSigma_{\bnu_1|\bnu_2}
\end{array}\right)
+\sigma_{\epsilon}^2 (\bK_1^T\bK_1)^{-1} \right).
\end{equation}
The $J_1$-dimensional vector $\mu_{\boldeta}$ and $J_1 \times J_1$ matrix $\Sigma_{\boldeta}$ are the mean and variance of $\boldeta(\btheta,\bY_1^R)$.  The $M$-dimensional vector $\boldsymbol{\mu}_{\bnu_1|\bnu_2}$ and $M\times M$ matrix $\bSigma_{\bnu_1|\bnu_2}$ are the conditional mean and variance of $\bnu_1$ given $\bnu_2$ which can be computed as
\begin{align*}
\boldsymbol{\mu}_{\bnu_1|\bnu_2}&= \frac{1}{\alpha_2^2} \mathbf{R}_\nu \bnu_2, \\
\bSigma_{\bnu_1|\bnu_2}&=  \alpha_1^2 \left( \mathbf{I}_M- \mathbf{R}_\nu  \mathbf{R}_\nu ^T \right).
\end{align*}

Using the likelihood function corresponding to this probability model and some standard prior specification for $\btheta$, $\alpha_1^2$ and $\sigma_{\epsilon}^2$ (see Section \ref{section:results} for details), we can infer the parameters via Markov chain Monte Carlo (MCMC). The computational cost for likelihood evaluation reduces from $\frac{1}{3}n^3$ to $\frac{1}{3} (J_1+M)^3$.

\section*{Appendix B. Detailed Description for the Posterior Density based on the Model Specification in Section \ref{subsubsectoin:RedDimCalib}}

The parameters that we estimate in the equations in \eqref{equation:obs1} and \eqref{equation:obs2} are the input parameter $\btheta$ (which is our main target),  the variance of the i.i.d observational errors for the grounding line positions $\sigma_{\epsilon}^2$, coefficients for the emulator term $\bpsi=\bpsi(\btheta,\bY_2^R)$, the coefficients for the discrepancy term for the modern binary pattern $\bnu_2$, and the variances of $\bnu_1$ and $\bnu_2$, $\alpha_1^2$ and $\alpha_2^2$. In addition to these parameters we also re-estimate the sill parameters for the emulator $\boldeta$, $\boldsymbol{\kappa}_1=[\kappa_{1,1},\dots,\kappa_{1,J_1}]$ \citep[cf.][]{bayarri2007computer,bhat2013inferring,chang2013fast} to account for a possible scale mismatch between the computer model output $\bY_1$ and the observational data $\bZ_1$. However, we do not re-estimate the sill parameters for $\bpsi$, $\kappa_{2,1},\dots,\kappa_{2,J_2}$, since both $\bY_2$ and $\bZ_2$ are binary responses and hence a scaling issue is not likely to occur here; in fact we have found that re-estimating these parameters causes identifiability issues between the emulator term $\bK_{2,y} \bpsi(\btheta,\bY_2^R)$ and the discrepancy term $ \bK_{2,d} \bnu_2$. 

The posterior density can be written as

\begin{align*} \small
\pi(\btheta,\bpsi,\boldsymbol{\kappa}_1,\bnu_2,\alpha_1^2,\alpha_2^2,\sigma_{\epsilon}^2,\mathbf{R}_\nu|\bY_1^R,\bZ_1^R,\bY_2^R,\bZ_2)\propto & ~ L\left(\bZ_1^R|\bY_1^R,\btheta,\boldsymbol{\kappa}_1,\alpha_1^2,\sigma_{\epsilon}^2,\bnu_2, \mathbf{R}_\nu\right) \\ 
&\times f(\boldsymbol{\kappa}_1) f(\alpha_1^2) f(\sigma_{\epsilon}^2) f(\mathbf{R}_\nu)\\
&\times L\left( \bZ_2|\bpsi,\bnu_2\right)\\
& \times f\left(\bpsi|\btheta,\bY_2^R\right) f(\bnu_2|\alpha_2^2) f(\alpha_2^2 ),\\
& \times  f(\btheta)
\end{align*}
The likelihood function $L\left(\bZ_1^R|\bY_1^R,\btheta,\boldsymbol{\kappa}_1,\alpha_1^2,\sigma_{\epsilon}^2,\bnu_2, \mathbf{R}_\nu\right) $ is given by the probability model in \eqref{eqn:Z_R prob}. For $f(\boldsymbol{\kappa}_1)=f(\kappa_{1,1},\dots,\kappa_{1,J_1})$ we use independent inverse gamma priors with a shape parameter of 50 and scale parameters specified in a way that the modes of the densities coincide with the estimated values of $\kappa_{1,1},\dots,\kappa_{1,J_1}$ from the emulation stage. We assign a vague prior $IG(2,3)$ for $f\left(\alpha_1^2\right)$, $f\left(\alpha_2^2\right)$, and $f\left(\sigma_{\epsilon}^2\right)$, and a uniform prior for $f(\btheta)$ whose support is defined by the range of design points $\btheta_1,\dots,\btheta_p$. The likelihood function $L(\bZ_2|\bpsi,\bnu_2)$ is defined as 

\begin{align*}
L\left( \bZ_2|\bpsi,\bnu_2\right) &\propto \prod_{j=1}^n \left( \frac{\exp(\lambda_j)}{1+\exp(\lambda_j)} \right)^{Z_2(\bs_j)} \left( \frac{1}{1+\exp(\lambda_j)} \right)^{1-Z_2(\bs_j)},
\end{align*}
where $\lambda_j$ is the $j$th element of $\blambda$ in \eqref{equation:obs2}. The conditional density $f\left(\bpsi|\btheta,\bY_2^R\right)$ is given by the Gaussian process emulator  $\bpsi(\btheta,\bY_2^R)$. The conditional density $f\left( \bnu_2|\alpha_2^2 \right)$ is defined by the model $\bnu_2 \sim N(0,\alpha_2^2 \mathbf{I}_{L})$. The prior density $f(\mathbf{R}_\nu)$ is defined as 

\begin{equation*}
\prod_{i=1}^{M} \prod_{j=1}^{L} I(-1<\rho_{\nu,i,j}<1) \cdot I( \mathbf{I}_M- \mathbf{R}_\nu  \mathbf{R}_\nu ^T \mbox{ is positive definite}),
\end{equation*}
where $I(\cdot)$ is the indicator function and $\rho_{\nu,i,j}$ is the $(i,j)$th element of $\mathbf{R}_\nu$.

\bibliography{short,references}
\clearpage

\end{document}